\documentclass[%
 reprint,
superscriptaddress,
 amsmath,amssymb,
 aps,
]{revtex4-2}

\usepackage{graphicx}% Include figure files
\usepackage{dcolumn}% Align table columns on decimal point
\usepackage{bm}% bold math
\usepackage{bbm}
\usepackage{color}
\usepackage{hyperref}
\usepackage[dvipsnames]{xcolor}
\usepackage[version=4]{mhchem}
\usepackage{verbatim}
\usepackage{makecell}

\begin{document}
\preprint{APS/123-QED}
%\articletype{Research Article}
\title{Inverse design for robust inference in integrated computational spectrometry}
\author{Wenchao~Ma}
\affiliation{Department of Chemistry, Massachusetts Institute of Technology, Cambridge, MA 02139, USA}
\author{Rapha{\"e}l~Pestourie}
\affiliation{School of Computational Science and Engineering, Georgia Institute of Technology, Atlanta, GA 30332, USA}
\author{Zin~Lin}
\affiliation{Bradley Department of Electrical and Computer Engineering, Virginia Tech, Blacksburg, VA 24061, USA}
\author{Steven~G.~Johnson}
\email{stevenj@math.mit.edu}
\affiliation{Department of Mathematics, Massachusetts Institute of Technology, Cambridge, MA 02139, USA}

%\runningtitle{Running title}
\begin{abstract}
We propose an inverse-design approach for computational spectrometers in which the scattering media are topology-optimized to achieve better performance in inference of unknown spectra. Unlike traditional end-to-end approaches, our inverse design of the scattering media does not need a training set of spectra, a distribution of detector noise, or an inference algorithm. Our approach allows the selection of the inference algorithm to be decoupled from that of the scatterer. For smooth spectra, we additionally devise a regularized reconstruction algorithm based on Chebyshev interpolation, which yields higher accuracy compared with conventional methods in which the spectra are sampled at equally spaced frequencies or wavelengths with equal weights. Our approaches are numerically demonstrated via inverse design of integrated computational spectrometers and reconstruction of example spectra. The inverse-designed spectrometers exhibit significantly better performance in the presence of noise than their counterparts with random scatterers. Our method provides a useful complement to end-to-end co-design methods.

\end{abstract}

%\journalname{Nanophotonics}
%\dedication{}
%\journalyear{2025}
%\journalvolume{aop}
\maketitle

\section{Introduction}

Conventional computational spectrometry (Sec.~\ref{sec:spectrometer}) attempts to reconstruct the spectrum of input light by analysis of the light scattered through a complex medium (Fig.~\ref{sketch}), often a disordered medium, exploiting the fact that the recorded signal is a superposition of frequency-dependent scattering patterns~\cite{Yang2021,Guan2023,Xue2024}.  Although many different algorithms have been applied to this reconstruction~\cite{Xu2003,Chang2008,Redding2013NP,Redding2013,Wang2014,Kurokawa2011,Yang2012,Yang2013,Hang2010,Oliver2012,August2013,Wang2014cs,Liew2016,Kita2018,Chang2011,Chang2012,Yang2019}, usually by some form of optimization/regression problem, previous work typically takes the scattering medium itself as given, or perhaps selects from a small menu of randomized geometries~\cite{Redding2013NP}.  In this work, we address the question of whether a better scattering medium can be \emph{inverse-designed} for computational spectrometry, optimizing the medium itself over a vast number ($\approx 4\times10^5$) of parameters in order to maximize some measure of ``information throughput'' and/or robustness against noise for the subsequent computational inference.   Of course, given an arbitrarily large scattering volume and enough sensors, one can make computational inference easier simply by designing a prism/demultiplexer (where different wavelengths are designed to scatter to different sensor regions)~\cite{Kim2012,Wang2014,Hadibrata2021}, but the challenge is to obtain accurate reconstruction with a small scatterer (e.g.~integrated onto a chip~\cite{Li2022advances,Peters2025,Zhang2021,Redding2013NP,Velasco2013,HerreroBermello2017,Podmore2017,Kita2018,Hadibrata2021,Li2021,Tian2024}) and a few sensors (e.g.~a discrete set of output waveguides).

We demonstrate that it is possible to achieve order-of-magnitude improvement in reconstruction robustness against sensor noise, compared with the median performance of an ensemble of random scatterers, by inverse-designing the scatterer to improve inference.
Rather than ``end-to-end'' design where one co-optimizes inference and scattering to directly minimize reconstruction error~\cite{Sitzmann2018,Lin2021,Lin2022,Li2023,Arya2024}---which requires a training set of spectra and the assumptions on the distribution of detector noise and a reconstruction algorithm as well as various additional hyperparameters such as the mini-batch size and learning rate in stochastic optimization---we instead show that an efficient and interpretable alternative is to optimize a measure of the ``inference robustness'' of the scattering system given by a nuclear norm (also called trace norm)~\cite{Horn1991,Fazel2001} of the pseudo-inverse of a measurement matrix (relating input spectra to sensor readings).
Our nuclear-norm formulation simultaneously addresses two performance goals: different frequencies should scatter into very distinct sensor readings (leading to ``well-conditioned'' reconstruction), but the collection efficiency should also be high at all frequencies (related to improving signal-to-noise ratios).
We show that our approach is tractable for freeform topology optimization (TopOpt), where ``every pixel'' is a degree of freedom, demonstrated theoretically in an example two-dimensional (2d) system modeling an integrated-optics spectrometer.  Our example system employs a single dielectric waveguide as input, passes light with wavelengths $\lambda \in [1540,1560]$~nm through a scatterer smaller than $10\lambda$, and reconstructs a continuous spectrum using 12 output waveguides.
Our optimized structure (which behaves very unlike a prism) exhibits greatly improved robustness against noise for computational inference by least-squares (overdetermined) reconstruction, even though no explicit noise, training spectra, or reconstruction algorithm were used during the scatterer optimization itself. The fact that our objective function is directly related to the singular values of the measurement matrix allows us to immediately interpret \emph{why} this robustness occurs.
Another key challenge is relating discrete measurements to reconstructing a spectrum over a continuous frequency range.  For spectra that vary smoothly with frequency, we devise a regularized reconstruction algorithm that exploits this smoothness using Gauss--Legendre quadrature~\cite{Atkinson1991} and Chebyshev polynomial interpolation~\cite{Trefethen2019,Boyd2001}, yielding greater accuracy than typical methods in which the spectrum is sampled at equally spaced and equally weighted points.
We believe that our approach, which separates the design of an improved scatterer from the specifics of reconstruction, should enable efficient exploration of future computational spectrometry systems and algorithms (Sec.~\ref{sec:conclusion}), and related approaches may also be applicable to designing optics for computational imaging~\cite{Eils2003,Mait2018,Hu2024,Xiang2024,Zhang2022,Bacca2023} or other inference problems~\cite{Li2023}.

Inverse design employs large-scale optimization to maximize optical performance, measured by some figure of merit (FOM), over a huge number of geometric and/or material degrees of freedom (DOFs)~\cite{Molesky2018,Kang2024}. Specifically, in this work we employ topology optimization (TopOpt)~\cite{Jensen2010,Christiansen2021,Chen2024}, in which freeform geometries are optimized over ``every pixel'' of the structure, typically while imposing manufacturing constraints~\cite{Zhou2015,Chen2024}.   A key enabling factor of inverse design is that one can compute the derivatives (i.e., the gradient) of an FOM with respect to a huge number of DOFs (e.g. the material at every pixel) by a single additional ``adjoint'' solve of essentially the same (Maxwell) equations~\cite{Molesky2018,Kang2024,LalauKeraly2013}, which can then be used for gradient ascent and related optimization algorithms~\cite{Boyd2004,Nocedal2006,Nesterov2018}.  When the output of the optics is fed into computational inference, however, one should target a different FOM: instead of designing optics for a pre-determined wave-scattering operation (e.g.~a coupler), one wishes to minimize the error of the inference in the presence of noise.   The most direct approach to improving inference is  end-to-end co-design: given a training set of inputs and a distribution of detector noise, one minimizes the mean error in the final inference with respect to geometry and inference DOFs (backpropagating the gradient of the error through both the inference and the Maxwell solves), and this strategy has been recently applied to several problems in imaging and other applications~\cite{Sitzmann2018,Lin2021,Lin2022,Li2023,Arya2024}.  The end-to-end approach requires a training set, a noise distribution, and an inference model. Incorporating explicit noise and/or random training-set sampling leads towards stochastic-optimization algorithms~\cite{Murphy2022}, excluding many other optimization methods unless additional approximations are made. Some authors instead optimize a deterministic training-data--free proxy for inference robustness, such as measures of mutual information~\cite{Pinkard2025} or Fisher information~\cite{Kabuli2025}, which are related to our nuclear-norm formulation (Sec.~\ref{sec:obj}).  We describe an explicit comparison to end-to-end methods for our spectrometry example in Sec.~\ref{sec:robustness}.  Related work has instead sought theoretical upper bounds on the information throughput of optical systems, e.g.~in terms of channel capacity~\cite{Tang2001,Essiambre2010,Shtaif2022,Amaolo2025}, although these bounds do not identify specific practical designs.

\section{Computational spectrometer}\label{sec:spectrometer}
\subsection{Forward model}
\begin{figure}[ht]
\centering
%\fbox{\includegraphics[width=0.7\linewidth]{sketch}}
\includegraphics[width=1\linewidth]{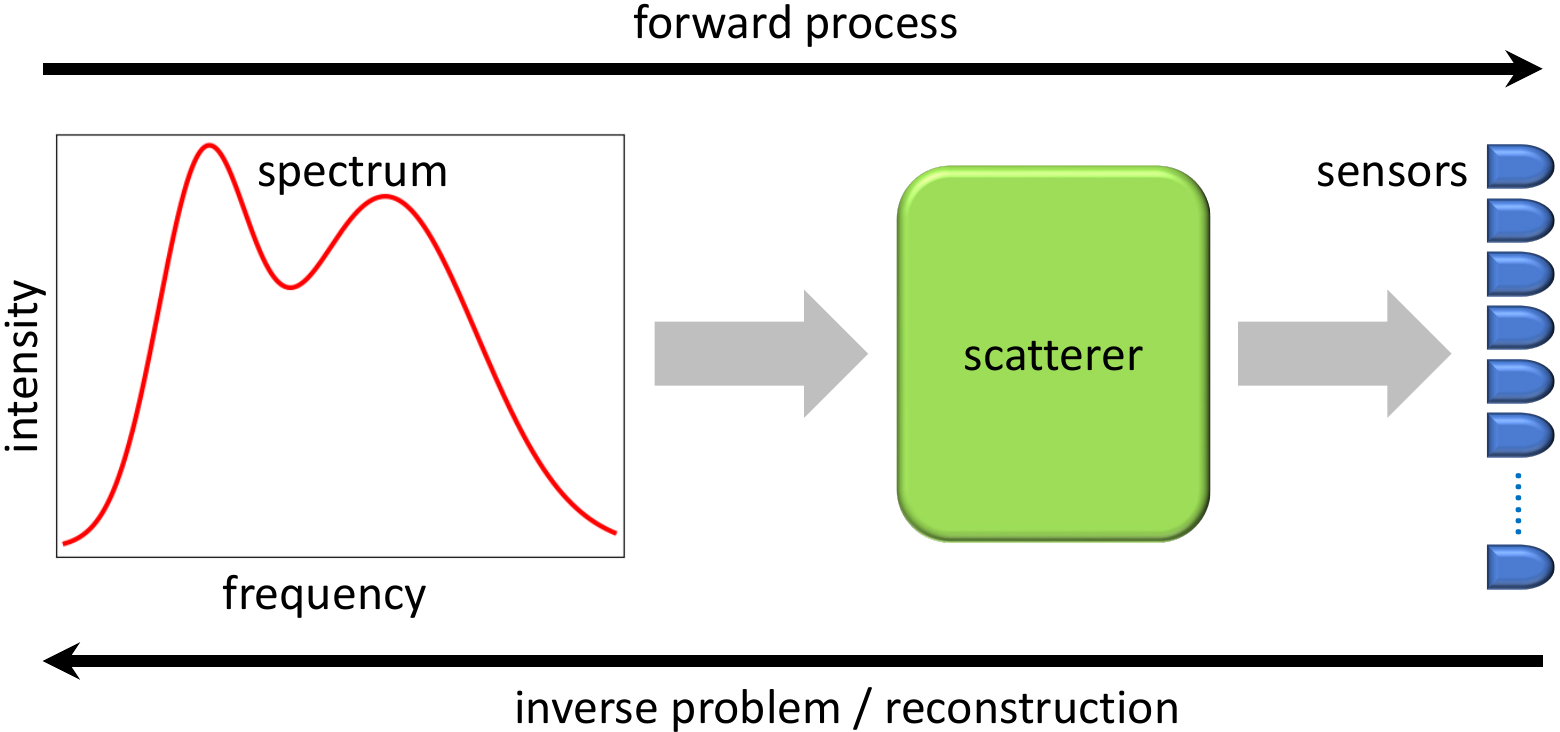}
\caption{Sketch of computational spectrometry: forward process and inverse problem.  In the forward process, input waves pass through a scatterer and form frequency-dependent patterns on sensors.  If this dependence is calibrated beforehand, one may reconstruct unknown spectra from the signals recorded by the sensors, which is an inverse problem.}
\label{sketch}
\end{figure}

In a conventional computational spectrometer, light passes through a scatterer and forms frequency-dependent patterns on sensors. (One can also employ multiple measurements through a reconfigurable scatterer on a single sensor~\cite{Velasco2013,HerreroBermello2017,Podmore2017,Kita2018}).  Output signals typically depend linearly on input power and the incoming light is incoherent at different frequencies.  For a spectrometer with a finite number of sensors as sketched in Fig.~\ref{sketch}, the power received by each sensor is an integrated power over a range of frequencies. In the presence of noise, the signal recorded at the $k$-th sensor can be written as
\begin{equation}\label{eq:sensor-k-intensity}
v_k = \int F_k(\omega) u(\omega)d\omega+\zeta_k,
\end{equation}
where $\omega$ denotes frequency, $u(\omega)$ denotes the input power at the frequency $\omega$ (i.e., the unknown spectrum to be determined below), $F_k(\omega)$ denotes the signal recorded at the $k$-th sensor due to unit input power at the frequency $\omega$, $\zeta_k$ denotes the noise on the $k$-th sensor, and $v_k$ denotes the power received by the $k$-th sensor. The function $F_k(\omega)$ describes the overall response the optical system between input and output ends, encapsulating the effects of various components, such as sensors, scattering media, filters, and substrates.
%In this work, different sensors are physically separate, but the mathematical formulation is the same for a single sensor used with a reconfigurable scatterer to make multiple distinct measurements~\cite{Velasco2013,HerreroBermello2017,Podmore2017,Kita2018}.

%\subsection{Discretization}
For a given spectrometer, estimating $u(\omega)$ (the spectrum) from all $v_k$ (the sensor measurements) and $F_k(\omega)$ (which can be determined from simulation or experimental calibration) requires decomposing $u(\omega)$ into a finite number of unknowns, such as expanding $u(\omega)$ in terms of a basis of functions with unknown coefficients, or discretizing frequency $\omega$ into a finite number of samples $\omega_j$ at which each $u(\omega_j)$ is unknown. Solving for these unknowns results in spectral reconstruction.

In a discretization scheme, the exact integral can be approximated by a weighted sum with weights $w_j$ at a set of discrete frequencies $\omega_j$ (according to a quadrature rule~\cite{Davis2007} and the domain of integration): 
\begin{equation}\label{eq:forward-matrix}
v_k\approx \sum_j 
w_jF_k(\omega_j)
u(\omega_j)+\zeta_k \quad
\Longleftrightarrow \quad
v\approx FWu+\zeta\, ,
\end{equation}
where on the right we have expressed the relation in matrix form: $u$, $v$, and $\zeta$ are column vectors of $u(\omega_j)$, $v_k$, and $\zeta_k$; $W={\rm diag}(w_1,w_2,\cdots)$ contains the weights. For example, the simplest quadrature rule is a Riemann sum with equally spaced frequencies $\omega_j = \omega_0 + j \Delta \omega$ and equal weights $w_j = \Delta \omega$~\cite{Davis2007}. 

%If all spectra under consideration are nonzero only at a set of discrete frequencies $\omega_j$ (which may not follow any quadrature rule), the relation between the input spectrum and recorded intensity is simply
%\begin{equation}%\label{eq:sensors-intensity}
%y = Fs+\eta\, .
%\end{equation}

One may also expand $u(\omega) \approx \sum_\ell b_\ell(\omega) c_\ell$ in some finite set of basis functions $b_\ell$ (e.g., polynomials such as the Chebyshev polynomials~\cite{Trefethen2019}, radial basis functions~\cite{Buhmann2003} such as Gaussians~\cite{Chang2011,Chang2012,Yang2019,Cheng2021}, etc.) and unknown coefficients $c_\ell$; in terms of this basis, the matrix equation and the spectrum vector in Eq.~(\ref{eq:forward-matrix}) can be expressed as
\begin{equation}\label{eq:template}
v\approx FWBc+\zeta\,,
\end{equation}
where $B$ is a matrix with elements $B_{j,\ell} = b_\ell(\omega_j)$ and $c$ is a column vector of $c_\ell$. If $B$ is an identity matrix, $c$ becomes $u$ and the matrix equation in Eq.~(\ref{eq:template}) becomes identical to that in Eq.~(\ref{eq:forward-matrix}).

\subsection{Reconstruction}
The key enabling factor for reconstruction is that the columns of $F$ are distinct: different frequencies yield different measured signals, allowing one to disentangle superpositions of multiple frequencies into their component amplitudes.
In this work, we consider only overdetermined problems, in which more measurements than unknowns are available. In terms of Eqs.~(\ref{eq:forward-matrix}) and (\ref{eq:template}), the vector $v$ contains more elements than $u$ or $c$.

To reconstruct an unknown continuous spectrum, a conventional approach is to seek a least-squares solution, corresponding to using the pseudo-inverse of the matrix $FW$ in Eq.~(\ref{eq:forward-matrix}):
\begin{equation}\label{eq:discrete-reconst}
\hat u=(FW)^+v,
\end{equation}
where $\hat u$ denotes the reconstructed spectrum and the superscript + denotes pseudo-inversion~\cite{Trefethen2022}, which becomes ordinary matrix inversion if $FW$ is a square matrix. The continuous spectrum can then be recovered via interpolation and extrapolation (or only interpolation if the discrete frequencies span the full range of interest).
%(The discrete set of frequencies in reconstruction may not necessarily be the same as those in optimization.)
One may also estimate the coefficients of basis functions in Eq.~(\ref{eq:template}) and then reconstruct the spectrum:
\begin{equation}\label{eq:template-reconst}
\hat c=(FWB)^+v,\qquad \hat u=B\hat c,
\end{equation}
without the need for an extra interpolation or extrapolation step. Although the spectrometry problem is underdetermined because intensities at infinitely many frequencies are reconstructed from a finite number of measurements, the basis functions impose a smoothness prior that allows the problem to become determined or overdetermined if the number of reconstructed coefficients does not exceed the number of measurements.

To reduce reconstruction error further for overdetermined problems in the presence of noise, Tikhonov regularization~\cite{Hansen1998} is typically employed. In terms of expansion with basis functions in Eqs.~(\ref{eq:template}) and (\ref{eq:template-reconst}), one can estimate the coefficients and spectra as
\begin{equation}\label{eq:c-argmin}
\hat c = \arg\min_{c}\left( \left\Vert FWBc-v\right\Vert_2^2+\alpha\left\Vert \sqrt{W}Bc\right\Vert_2^2\right),
\end{equation}
where $\alpha$ is a regularization coefficient. %and $\left\Vert B\mu\right\Vert_2^2$ is understood as $(B\mu)^\top WB\mu$, if $B\mu$ is discretized at those quadrature nodes.
The estimated coefficients $\hat c$ can be analytically solved and an unknown spectrum can then be reconstructed:
\begin{equation}\label{eq:c-reconstr}
\begin{aligned}
\hat c &= \left[B^\top(WF^\top FW+\alpha W)B\right]^{-1}(FWB)^\top v,\\
\hat u &= B\hat c.
\end{aligned}
\end{equation}
%More complicated techniques, such as neural networks~\cite{Aggarwal2018}, may also be applied, especially to underdetermined reconstruction problems, in which more unknowns than measurements need to be estimated~\cite{Hansen1998,Hansen2013}.[Ref for research]
Other computational spectrometry work, especially in the underdetermined case, has also explored other methods such as sparsifying $L_1$  regularization~\cite{Podmore2017,Kita2018,Oliver2012,August2013,Wang2014cs,Liew2016,Foucart2013,Hastie2015} and even neural networks~\cite{Brown2021,Yang2023,Liao2024,Wang2024,Wang2024oe,Liang2024}.
%The above reconstruction approaches rely on the assumption that the influence of noise $\zeta$ is insignificant, which often requires the reconstruction process is well conditioned. Although regularization and some more complicated techniques may be applied~\cite{Hansen1998,Hansen2013}, robustness against noise is still favorable. 

\section{Inverse-design method}
\subsection{Performance metrics}
In this work, we focus on two performance metrics: robustness against noise on sensors and collection efficiency of signals (which is related to signal-to-noise ratios).

The robustness is related to the ratio of the relative error in the reconstructed spectrum, namely $\|\hat u-u\|/\|u\|$, to the relative error in the signal, namely $\|\zeta\|/\|v\|$. This ratio is bounded above by the condition number of the matrix $F\sqrt{W}$~\cite{Trefethen2022}, up to the discretization error in $\omega$.
To see how robustness connects with $F\sqrt{W}$, we first consider the difference between the true and reconstructed spectra when Eq.~(\ref{eq:forward-matrix}) is used:
\begin{equation}
\hat u-u \approx (FW)^{+}\zeta=W^{-1}F^+\zeta,
\end{equation}
where the last equality relies on the assumption that $F$ has linearly independent columns, which requires that frequencies are no more than measurements.
We quantify the error of reconstruction as
\begin{equation}\label{eq:reconstr-error-noreg}
\begin{aligned}
&\sqrt{\int \left[\hat u(\omega)-u(\omega)\right]^2d\omega}
\approx \sqrt{\sum_i \left[\hat u(\omega_i)-u(\omega_i)\right]^2w_i} \\
&=\sqrt{(\hat u-u)^{\top}W(\hat u-u)}=\sqrt{(F^+\zeta)^{\top}W^{-1}(F^+\zeta)}\\
&=\left\Vert \left(F\sqrt{W}\right)^+\zeta\right\Vert_2,
\end{aligned}
\end{equation}
with $\top$ denoting matrix transposition, and $\|\cdot\|_2$ denoting the $L_2$ norm. Likewise, we have
\begin{equation}%\label{eq:reconstr-error-noreg}
\sqrt{\int u(\omega)^2d\omega}
\approx \sqrt{\sum_i u(\omega_i)^2w_i}
=\left\Vert\sqrt{W}u\right\Vert_2.
\end{equation}
Therefore, the ratio of the relative error in the reconstructed spectrum to the
relative error in the signal is
\begin{equation}\label{eq:error-ratio}
\begin{aligned}
&\frac{\sqrt{\int \left[\hat u(\omega)-u(\omega)\right]^2d\omega}}{\sqrt{\int u(\omega)^2d\omega}}\Bigg/\frac{\|\zeta\|_2}{\|FWu\|_2}\\
&\approx\frac{\left\Vert{\left(F\sqrt{W}\right)}^+\zeta\right\Vert_2}{\left\Vert\sqrt{W}u\right\Vert_2}\Bigg/\frac{\|\zeta\|_2}{\|FWu\|_2}\\
&=\frac{\left\Vert{\left(F\sqrt{W}\right)}^+\zeta\right\Vert_2}{\|\zeta\|_2}\frac{\|F\sqrt{W}\sqrt{W}u\|_2}{\left\Vert\sqrt{W}u\right\Vert_2}.
\end{aligned}
\end{equation}
As a standard approach is then to maximize over $\zeta$ and $u$~\cite{Trefethen2022}, in which case the two factors after the equality become the $L_2$-induced matrix norms $\|{(F\sqrt{W})}^+\|_2$ and $\|{F\sqrt{W}}\|_2$, the product of which is the condition number of the matrix $F\sqrt{W}$ defined with the same norm, which is also the ratio between the maximum and minimum singular values~\cite{Trefethen2022}:
\begin{equation}\label{eq:cond}
\kappa\left(F\sqrt{W}\right)=\left\Vert{F\sqrt{W}}\right\Vert_2\left\Vert{\left(F\sqrt{W}\right)}^+\right\Vert_2=\frac{\sigma_{\max}(F\sqrt{W})}{\sigma_{\min}(F\sqrt{W})}.
\end{equation}
As an upper bound of the ratio of relative errors in Eq.~(\ref{eq:error-ratio}), the condition number can be regarded as a performance metric of a computational spectrometer.

A low condition number, however, is not sufficient for good performance: if all of the singular values are small, that would signify low collection efficiency, even if the $\sigma$ ratios are close to~1. Large signal intensities are beneficial for robust inference in the presence of noise components that do not increase (e.g., background light) or increase sublinearly with signal intensities (e.g., shot noise, which increases as $\sim \sqrt{\text{intensity}}$~\cite{Haus2000}). Therefore, the collection efficiency, e.g., transmittance in a transmission spectrometer, should also be a performance metric.  Here we report a straightforward way to incorporate both criteria (high collection efficiency and low condition number) into a single differentiable figure of merit, described in Sec.~\ref{sec:obj} below.

This efficiency can be considered as the ratio of the power of signals recorded by sensors (in the limit of no noise) to the power of an input spectrum with uniform intensity across the operational frequency range:
\begin{equation}\label{eq:efficiency}
\eta=\int \sum_k F_k(\omega)\approx\sum_{j,k} w_j F_k(\omega_j) \,.
\end{equation}

\subsection{Objective function}\label{sec:obj}
\begin{figure*}[ht]
\centering
%\fbox{\includegraphics[width=0.7\linewidth]{sketch}}
\includegraphics[width=0.75\linewidth]{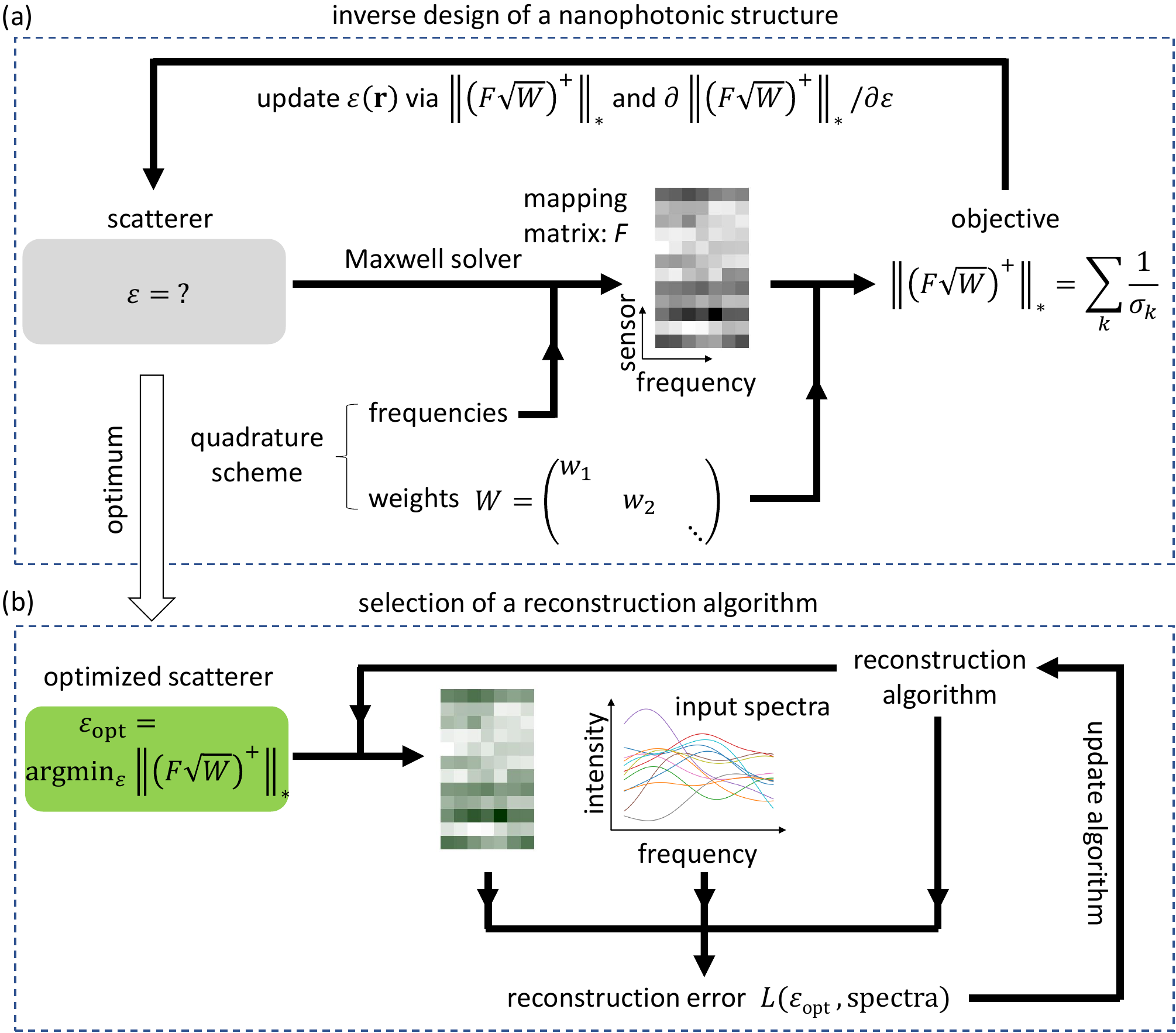}
\caption{Framework of design methods. (a) Inverse design of a nanophotonic structure. The frequencies at which the spectral--spatial mapping matrix (also called the measurement matrix in this paper) is computed are determined by the frequency range and the quadrature scheme. This mapping matrix and the quadrature weights determine the objective function.
(b) Selection of a reconstruction algorithm. After the optimized scatterer is obtained, given prior knowledge of input spectra, one can select a reconstruction algorithm based on reconstruction error.}
\label{flow}
\end{figure*}

A spectrometer with good performance should be robust against noise on sensors while having acceptable collection efficiency.
From Eqs.~(\ref{eq:error-ratio}--\ref{eq:cond}) the condition number $\kappa \ge 1$ should be low, whereas the efficiency $\eta$ from Eq.~(\ref{eq:efficiency}) should be high. To simultaneously account for both performance metrics in gradient-based optimization, instead of formulating a multi-objective optimization problem, we introduce a single figure of merit (FOM) to be \emph{minimized}:
\begin{equation}\label{eq:obj-singvals}
\left\Vert (F\sqrt{W})^+\right\Vert_*={\rm tr}\,\left[(WF^{\top}F)^{-1/2}\right]=\sum_j\frac{1}{\sigma_j}.
\end{equation}
where $\|\cdot\|_*$ means taking the nuclear norm (also called trace norm)~\cite{Horn1991,Fazel2001} and $\sigma_j$ denotes each singular value of $F\sqrt{W}$.  [Using the trace expression here facilitates the computation of gradients and makes this objective compatible with automatic differentiation. The flow chart is illustrated in Fig.~\ref{flow}(a).]   Clearly, minimizing this FOM tries to make all the singular values $\sigma_j$ larger, which implies higher collection efficiency for all relevant input spectra. On the other hand, the collection efficiency is bounded above by 100\%, indicating the existence of upper bounds on these singular values. Therefore, making them larger tends to decrease the \emph{spread} of singular values. In particular, the FOM is dominated by and has the sharpest dependence on the smallest $\sigma_j$, implying that this singular value is likely to enjoy the most relative increase as the FOM is minimized. Consequently, both a low condition number and a high collection efficiency are encouraged. We observe these two effects below in Fig.~\ref{optimize}.

In our approach, the optimization can evaluate transmission at relatively few frequencies (few columns of $F$), corresponding to a low-order quadrature scheme. In contrast, modeling the reconstruction step requires one to explicitly compute the measured noisy-sensor readings $v$ accurately for training/test spectra, involving a denser set of frequencies. An end-to-end approach which directly incorporates reconstruction into optimization can thus be more computationally costly.  Furthermore, the use of a deterministic objective function in our approach allows a wide range of optimization algorithms to be employed, including algorithms that support nonlinear constraints, such as the manufacturing constraints that are almost always required in practical TopOpt~\cite{Lazarov2016,Cool2025}. Explicit random noise and sampling of training data in end-to-end methods implies a smaller menu of stochastic-optimization methods~\cite{Murphy2022}, which have more limited options for nonlinear constraints~\cite{Lu2021}.

Many other objective functions may have similar effects, such as ${\rm tr}\left[(WF^{\top}F)^{-1}\right]=\sum_j{\sigma_j}^{-2}$ which is implied by the Fisher information~\cite{Kienesberger2025}, and $-\ln\det (WF^{\top}F)=-2\sum_j\ln\sigma_j$~\cite{Cover2006}. In our test example discussed in Sec.~\ref{sec:results}, the former objective function yields an optimized structure with performance comparable to that of the design from Eq.~(\ref{eq:obj-singvals}), while the log-determinant objective leads to much worse performance. Further variations of these objective functions could be explored, such as ${\rm tr}\left\{[(WF^{\top}F)^{-1}]^{n/2}\right\}=\sum_j{\sigma_j}^{-n}$ with $n>0$ (apart from $n=1$ or 2 that result in the objective functions mentioned earlier). One may also separately design a target matrix for $F$ (which may not be physically attainable) and then optimize the structure to minimize the distance of its actual $F$ to this target~\cite{Yu2025}, but this approach may over-constrain the design by not allowing it the freedom to discover the best attainable $F$.

In this work, we choose the discrete frequencies $\omega_j$ and weights $w_j$ according to a Gauss–-Legendre quadrature rule, in order to maximize integration accuracy for a given number of points~\cite{Atkinson1991}. This determines the matrix $W$.  Below, we used a 7-point quadrature rule, appropriate for the smooth example spectra considered in our tests, but in general this choice will depend on the system of interest.  Enlarging the number of quadrature points increases the computational cost, although this is somewhat ameliorated by our use of a hybrid time/frequency-domain scheme that computes all frequencies simultaneously~\cite{Hammond2022}.

\subsection{Topology optimization}
For inverse design, we adopt density-based topology optimization, in which the design region is meshed and a ``density'' related to the permittivity at each pixel is a parameter to be optimized~\cite{Jensen2010}. This density $\rho(\mathbf{x})$ is defined on the design region and ranges in $[0,1]$. Before computing the permittivity, the density is blurred and then projected. The blurring operation can be described as convolution with a filter: $\tilde{\rho}({\bf x})=w\ast\rho$, where $\tilde{\rho}$ is the density after blurring, and we choose the filter $w$ as a conic filter, the radius of which is related to the minimum lengthscale of the design pattern~\cite{Zhou2015}. After blurring, a projection operation is performed to compute an almost-everywhere binary density $\hat{\tilde{\rho}}$, with a hyperparameter $\beta$ representing the binarization strength~\cite{Hammond2025}. The permittivity in the design region is then
\begin{equation}%\label{eq:template}
\epsilon=\epsilon_{\min}+(\epsilon_{\max}-\epsilon_{\min})\hat{\tilde{\rho}},
\end{equation}
where $\epsilon_{\min}$ and $\epsilon_{\max}$ are the minimum and maximum permittivities in the design region.
In optimization, one usually starts with small $\beta$ and gradually
increases it, so that the structure becomes binarized. In this work, we used $\beta=$ 2, 4, 8, 16, 32, and $\infty$, each of which spanned a number of iterations. After these iterations, minimum lengthscale constraints were imposed along with $\beta=\infty$ to prevent too small geometric features in the final design~\cite{Arrieta2025}.

During conventional density-based topology optimization, the structural parameters are updated by gradient-based optimization algorithms, in particular the CSSA algorithm (conservative convex separable approximation) with either the method of moving asymptotes (CCSA-MMA) or a quadratic penalty (CCSA-Q)~\cite{Svanberg2002}. In this work, we adopted CSSA-MMA before imposing lengthscale constraints and then CCSA-Q during the final set of iterations with lengthscale constraints. Both algorithms are implemented in NLopt, a free and open-source software package~\cite{Johnson-nlopt}. The gradient of the objective [the trace expression in Eq.~(\ref{eq:obj-singvals})] with respect to structural parameters can be rapidly obtained from an adjoint method, which consists of two simulations: the forward simulation of the original problem, and the adjoint simulation in which the adjoint sources related to the output instead of the input are placed. All electromagnetic simulations in optimization and verification were performed with a free and open-source implementation of the finite-difference time-domain (FDTD) method~\cite{Oskooi2010} and the inverse design was performed with its hybrid time/frequency-domain adjoint module~\cite{Hammond2022}.

\section{Results and discussions}\label{sec:results}
\subsection{Example structure}

\begin{figure*}[ht]
\centering
%\fbox{\includegraphics[width=0.7\linewidth]{sketch}}
\includegraphics[width=0.6\linewidth]{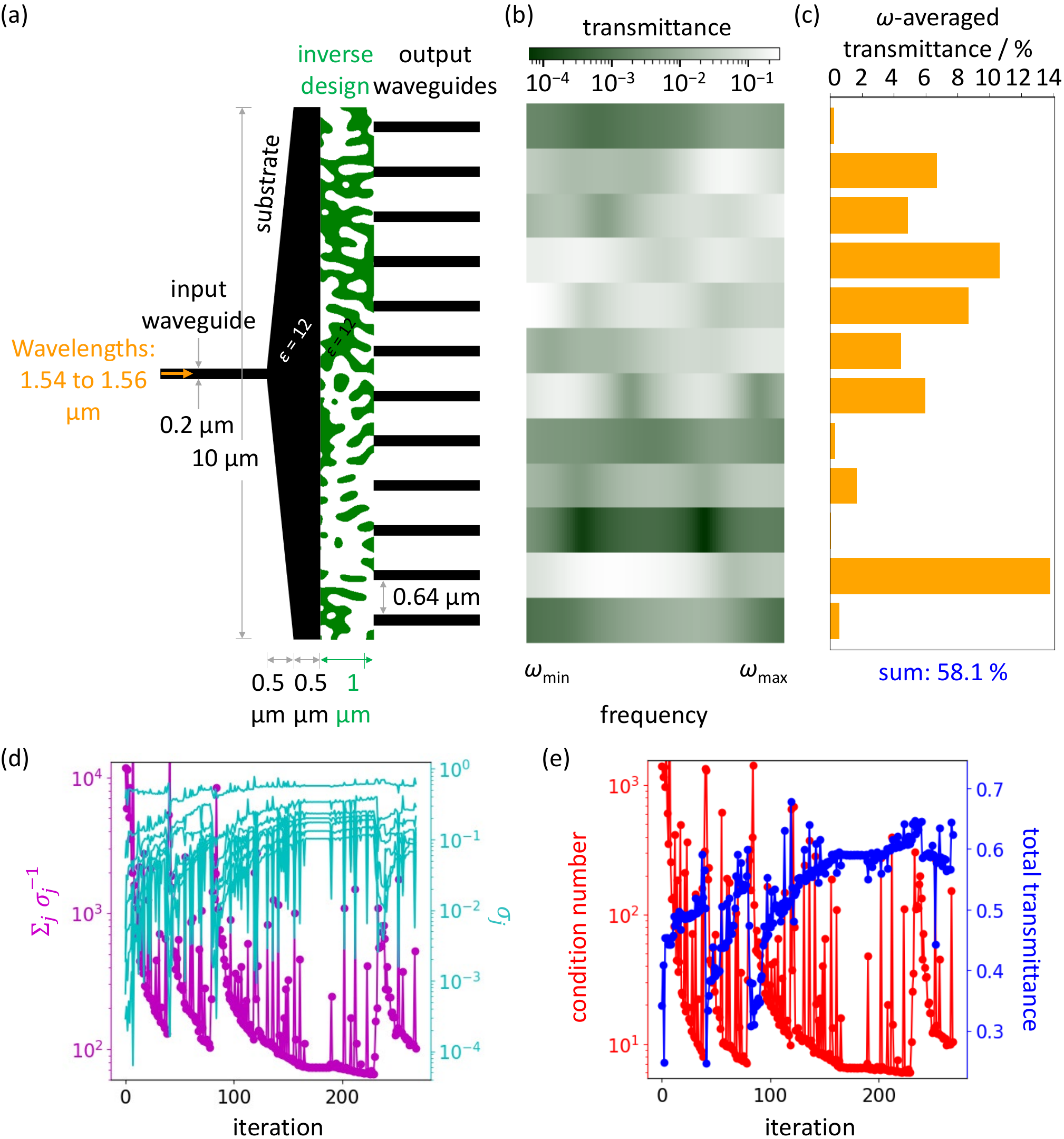}
\caption{Inverse design of an integrated spectrometer. (a) Structure of the spectrometer. This 2d device consists of an input waveguide, a wedge structure, a design region, and twelve output waveguides, with the solid material having a relative permittivity $\approx12$. The width of all the waveguides is 0.2 {\textmu}m. Adjacent output waveguides are separated by 0.64 {\textmu}m.
(b) Transmittance of the optimized spectrometer at each output waveguide across the frequency range of interest. (c) Frequency-averaged transmittance of the optimized spectrometer at each output waveguide. The total transmittance is 58.1\%. (d) Objective function ($\sum_j\sigma_j^{-1}$) and singular values of $F\sqrt{W}$ during optimization. (e) Condition number and total transmittance during optimization, computed from the $12\times7$ spectral--spatial mapping matrix.}\label{optimize}
\end{figure*}

Here, we demonstrate our methods on a simple two-dimensional (2d, $xy$) example of an integrated spectrometer.
As Fig.~\ref{optimize}(a) shows, the structure consists of an input waveguide, a wedge region, a design region, and twelve output waveguides, where the solid material has a relative permittivity of $3.48^2$ ($\approx12$, like silicon). (An alternative example system, in which the output waveguides are replaced by a uniform medium and far-field sensors, is given in Sec.~S5 in Supporting Information.) All waveguides have a width of 0.2 {\textmu}m and the separation between output waveguides is 0.64 {\textmu}m. Incoming waves at wavelengths 1.54 to 1.56 {\textmu}m with out-of-plane ($E_z$) polarization enter the wedge region from the input waveguide and undergo multiple scattering in the design region, the size of which is 10 {\textmu}m $\times$ 1 {\textmu}m. The scattering process, which is frequency-dependent, results in different output patterns at different frequencies, as Fig.~\ref{optimize}(b) shows. The input/output waveguides are single-mode in this wavelength range. We selected seven frequencies according to Gauss--Legendre quadrature of $\int_{1.54}^{1.56}d\lambda$ for performing inverse design, so the size of the spectral--spatial mapping matrix is $12\times7$.

%These two performance metrics are improved as the objective function is minimized.

\subsection{Inverse design}
%We performed inverse design with topology optimization, using the CCSA-MMA algorithm before imposing lengthscale constraints and the CCSAQ algorithm during the final bunch of iterations with lengthscale constraints~\cite{Svanberg2006}.
During minimization of Eq.~(\ref{eq:obj-singvals}), the smallest singular value increases most significantly, as Fig.~\ref{optimize}(d) shows. Meanwhile, as Fig.~\ref{optimize}(e) shows, the condition number of $F\sqrt{W}$ decreases from $>1000$ to $<100$, and the collection efficiency, which is the transmittance for this structure, increases from $30\sim40\%$ to approximately $60\%$. This high transmittance is unevenly distributed across the output waveguides, 7 of which dominate signal collection while 5 of which collect low portions of light, as shown in Figs.~\ref{optimize}(b) and (c). This behavior can be explained by the choice of 7 frequencies for optimization, which does not need to make full use of the 12 output channels. As the green-white region in Fig.~\ref{optimize}(a) illustrates, in the optimized design, a few high-transmittance output waveguides are not even connected to the high-index medium (silicon). This feature, although possible in 2d, would probably be absent in 3d in the presence of out-of-plane scattering.  (Note that the convergence history in Fig.~\ref{optimize}(d) is non-monotonic because we are also showing the ``inner'' iterations of the CCSA algorithm, in which it aggressively takes too large a step and backtracks by increasing a penalty~\cite{Svanberg2002}; there is also a discontinuity each time $\beta$ is increased.)

\begin{figure*}[ht]
\centering
%\fbox{\includegraphics[width=0.7\linewidth]{sketch}}
\includegraphics[width=0.55\linewidth]{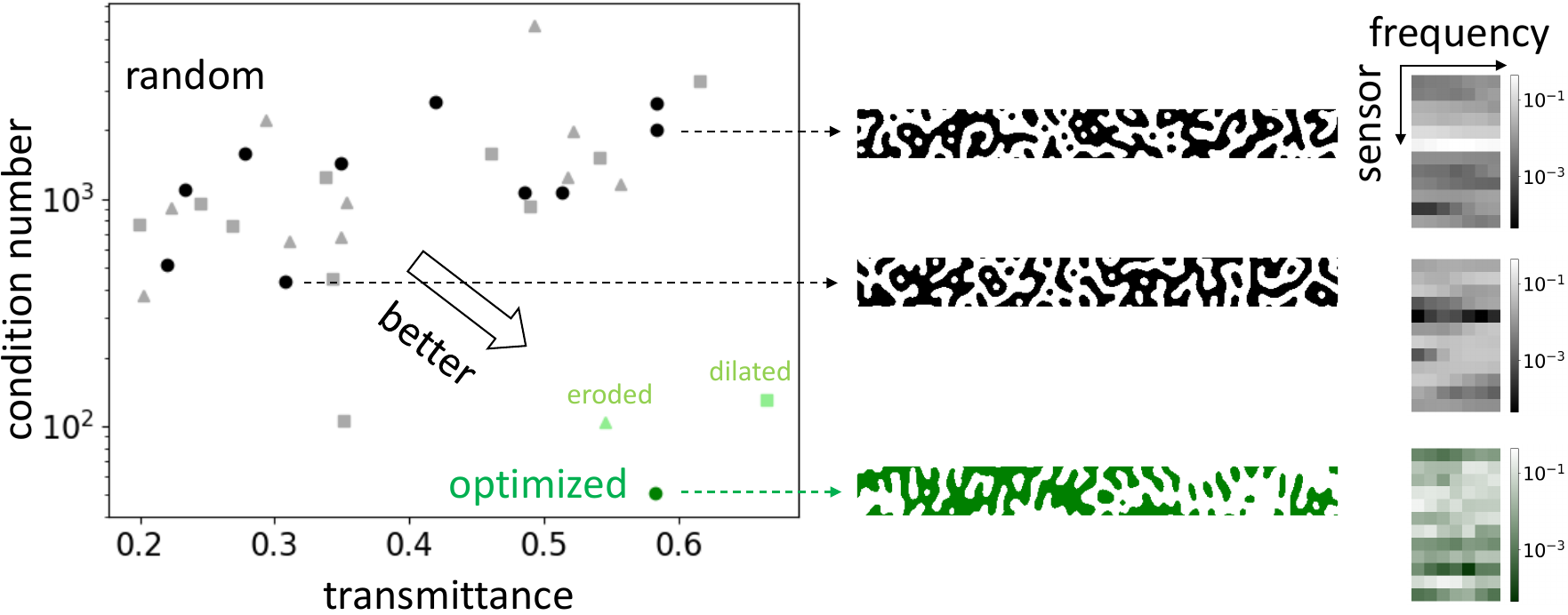}
\caption{Comparison of performances of random and optimized structures. The horizontal and vertical axes represent the total transmittance and the condition number, respectively. The bottom right area in this coordinate system is associated with lower condition numbers, higher collection efficiencies, and hence better performance. The round green and black dots correspond to the optimized and random structures with the same minimum lengthscale, while the square/triangular symbols with light colors correspond to structures in which solid regions are dilated/eroded by 10 nm. The optimized structure and two random structures are illustrated on the middle panel. Their corresponding $12\times7$ spectral--spatial mapping matrices are on the right panel.}\label{compare}
\end{figure*}

The optimized design pattern has a minimum lengthscale of 80~nm, measured by a free and open-source tool based on morphological transformations~\cite{Chen2024}. In comparison with some randomly generated structures with the same minimum lengthscale, the optimized structure clearly has better performance in its combination of low condition number and high transmittance, as Fig.~\ref{compare} shows. The performance metrics were evaluated using the 7 frequencies of the Gauss--Legendre nodes. To estimate the influence of fabrication imperfections~\cite{Siew2021,Boning2022}, we also simulated the morphological dilation and erosion by 10 nm of the structures, as also illustrated in Fig.~\ref{compare}: The condition number worsens by a factor of~2 to~3., and is still far better than typical random structures. It is also known that a low condition number makes the resulting inference error robust to perturbations in both the signal vector~$v$ (e.g., measurement noise) and the matrix~$F$ (e.g., manufacturing error)~\cite{Trefethen2022}. 
(Additional simulations of fabrication imperfections are shown in Sec.~S2 in Supporting Information.) As depicted in the right panel of Fig.~\ref{compare}, the order-of-magnitude smaller condition number corresponds to more-distinct columns in the $12\times7$ spectral--spatial mapping matrix of the optimized design. As we show explicitly below, this translates to greater robustness to noise in the reconstructed spectrum.

%As Fig.~\ref{optimize}(c) illustrates, the optimized spectral--spatial mapping matrix shows obviously different signal patterns at different frequencies, facilitating relatively faithful reconstruction of an unknown spectrum from a given signal pattern. 

%\subsection{Comparison with random structures}

\subsection{Reconstruction and impact of noise}
\label{sec:robustness}

\begin{figure}[h!]
\centering
%\fbox{\includegraphics[width=0.7\linewidth]{sketch}}
\includegraphics[width=1\linewidth]{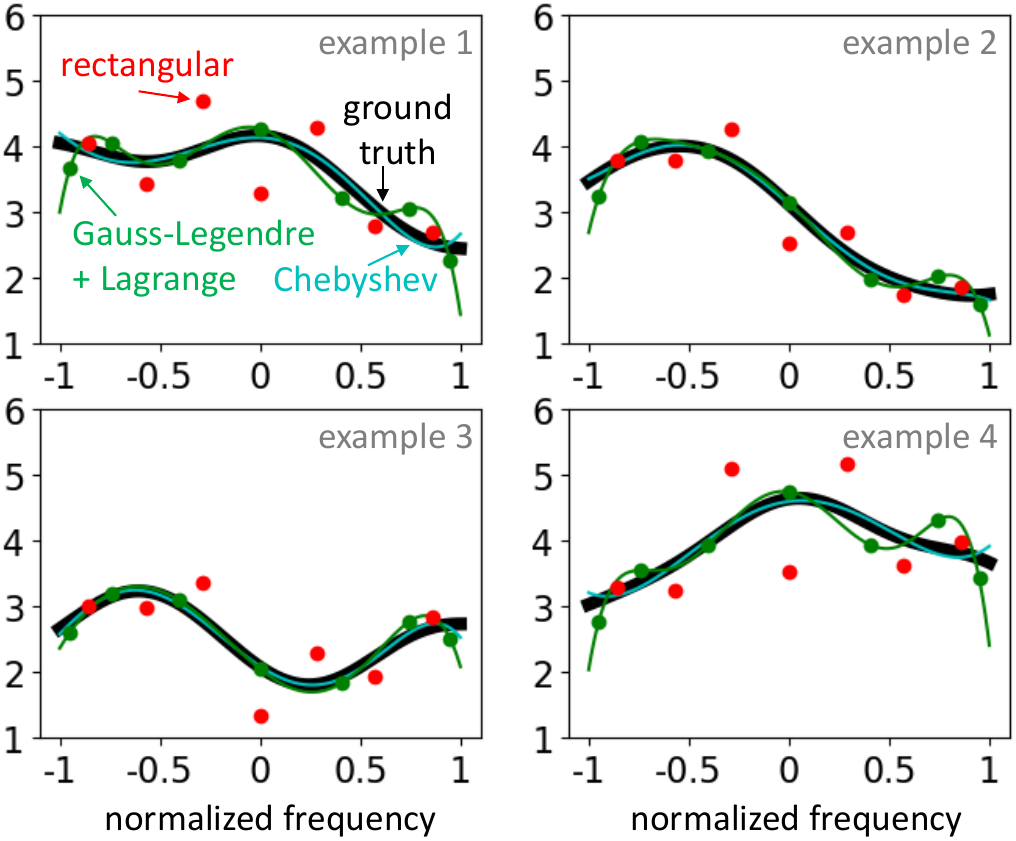}
\caption{Spectral reconstruction with four examples. The ground truth is plotted as thick black curves. The red dots represent reconstructed spectra at 7 equally spaced frequencies, chosen according the rectangular rule. The green dots represent reconstructed spectra at 7 frequencies chosen according to Gauss–Legendre nodes, while their Lagrange interpolating polynomials are plotted as green curves. The cyan curves represent reconstructed spectra as a linear combination of the first 7 Chebyshev polynomials of the first kind.}\label{reconstruct-7}
\end{figure}

Once the permittivity pattern is optimized, the first phase of design, as sketched in Fig.~\ref{flow}(b), is finished.
In this section, we turn to the second phase as sketched in Fig.~\ref{flow}(c), in which reconstruction algorithms are evaluated. Our sample spectra were randomly generated according to
\begin{equation}\label{eq:spectra}
\langle u(\omega_1)u(\omega_2)\rangle=\exp\left[-\frac{(\omega_1-\omega_2)^2}{\omega_{\rm corr}^2}\right],~~~~~~\langle u(\omega)\rangle=u_{\rm avg},
\end{equation}
where the correlation $\omega_{\rm corr}$ is chosen as $40\%$ of the frequency range and the average intensity $u_{\rm avg}$ is chosen as 3. Such spectra have smooth profiles and are usually positive everywhere. Four examples are shown as black curves in Fig.~\ref{reconstruct-7}. We assume independently and identically distributed (i.i.d.) noise on each sensor, obeying a normal distribution with zero mean and standard deviation proportional to the signal intensity on that sensor:
\begin{equation}\label{eq:noise}
\zeta_k\sim \mathcal{N}(\mu,\sigma_k^2),~~~~~~\mu=0,~~~~~~\sigma_k=qv_k,
\end{equation}
where $q$ describes the relative level of noise. To emulate the forward process, one needs to compute the signals faithfully. Here, we applied Eq.~(\ref{eq:forward-matrix}) using Gauss–Legendre quadrature with 101 points. These densely packed quadrature nodes in the frequency range of interest allows the signals to be computed accurately. On the other hand, the frequencies used for reconstruction hinge on the choices of reconstruction algorithms and may not be the same as those in computing the forward process.

As a natural choice, as described by Eq.~(\ref{eq:discrete-reconst}), one may reconstruct unknown spectra at a discrete set of frequencies, in particular, equally spaced frequencies. To avoid an underdetermined inverse problem, the number of such discrete frequencies should not exceed the number of sensors, which is 12 in our case. For example, let us consider 7 equally spaced frequencies chosen as the midpoints of intervals in the rectangular rule. Even in the absence of sensor noise $\zeta$, the reconstructed spectral intensities at these frequencies deviate from the true spectra, as shown by the blue dots in Fig.~\ref{reconstruct-7}(a), due to discretization error. Reconstruction at Gauss--Legendre nodes suffers from smaller error, as depicted by the green dots. One can then perform Lagrange interpolation and extrapolation to reconstruct continuous spectra, represented by the green curves.
%Although the reconstructed spectral intensities at these frequencies display the same trend as the ground truth, their mismatch is obvious even in the absence of sensor noise $\zeta$, because $F\sqrt{W}$ for reconstruction is only an approximation of the real mapping relation between spectra and signals. Interpolation and extrapolation from the reconstructed spectral intensities at these frequencies lead to further mismatch at some places, especially around the end points of the frequency range, as shown in Fig.~\ref{reconst}(a).

%also perform spectral reconstruction via interpolation and extrapolation from equally spaced frequencies that are chosen as the midpoints of intervals in the rectangular rule. With seven points, the spectral intensities at these frequencies are shown by the blue dots in Figs.~\ref{fig4}(a) while the blue curves pass through them illustrate the reconstructed continuous spectra, which obviously deviate from the ground truth even in the absence of noise.

Alternatively, one may also reconstruct unknown spectra as a linear combination of basis functions, as described by Eq.~(\ref{eq:template-reconst}). To avoid an underdetermined inverse problem, the number of basis functions should not exceed the number of sensors, while the number of frequencies is unlimited. For example, let us use Chebyshev polynomials of the first kind as the basis functions~\cite{Trefethen2019,Boyd2001}. Accordingly, $B$ in Eq.~(\ref{eq:template-reconst}) is a Chebyshev--Vandermonde matrix with rows and columns corresponding to different frequencies and different Chebyshev polynomials, respectively.  To make discretization error negligible, we used 101 frequencies located at the Gauss–Legendre quadrature nodes, which implies that $W$ is a $101\times101$ matrix. With 7 (zeroth- to sixth-order) such polynomials and the same densely packed frequencies as the forward model, in the absence of sensor noise $\zeta$, the reconstructed spectra, shown as the red curves in Fig.~\ref{reconstruct-7}(a), closely match the ground truth.

%We first consider spectral reconstruction using Chebyshev polynomials according to Eq.~(\ref{eq:template-reconst}), where $\hat c$ is the reconstructed coefficients of Chebyshev polynomials and $B$ is a Chebyshev-Vandermonde matrix with rows and columns corresponding to different frequencies and different Chebyshev polynomials, respectively. For example, with seven Chebyshev polynomials, in the absence of sensor noise $\zeta$, the reconstructed spectra, shown as the red curves in Figs.~\ref{fig4}(a), closely match the ground truth.

\begin{figure}[h!]
\centering
%\fbox{\includegraphics[width=0.7\linewidth]{sketch}}
\includegraphics[width=1\linewidth]{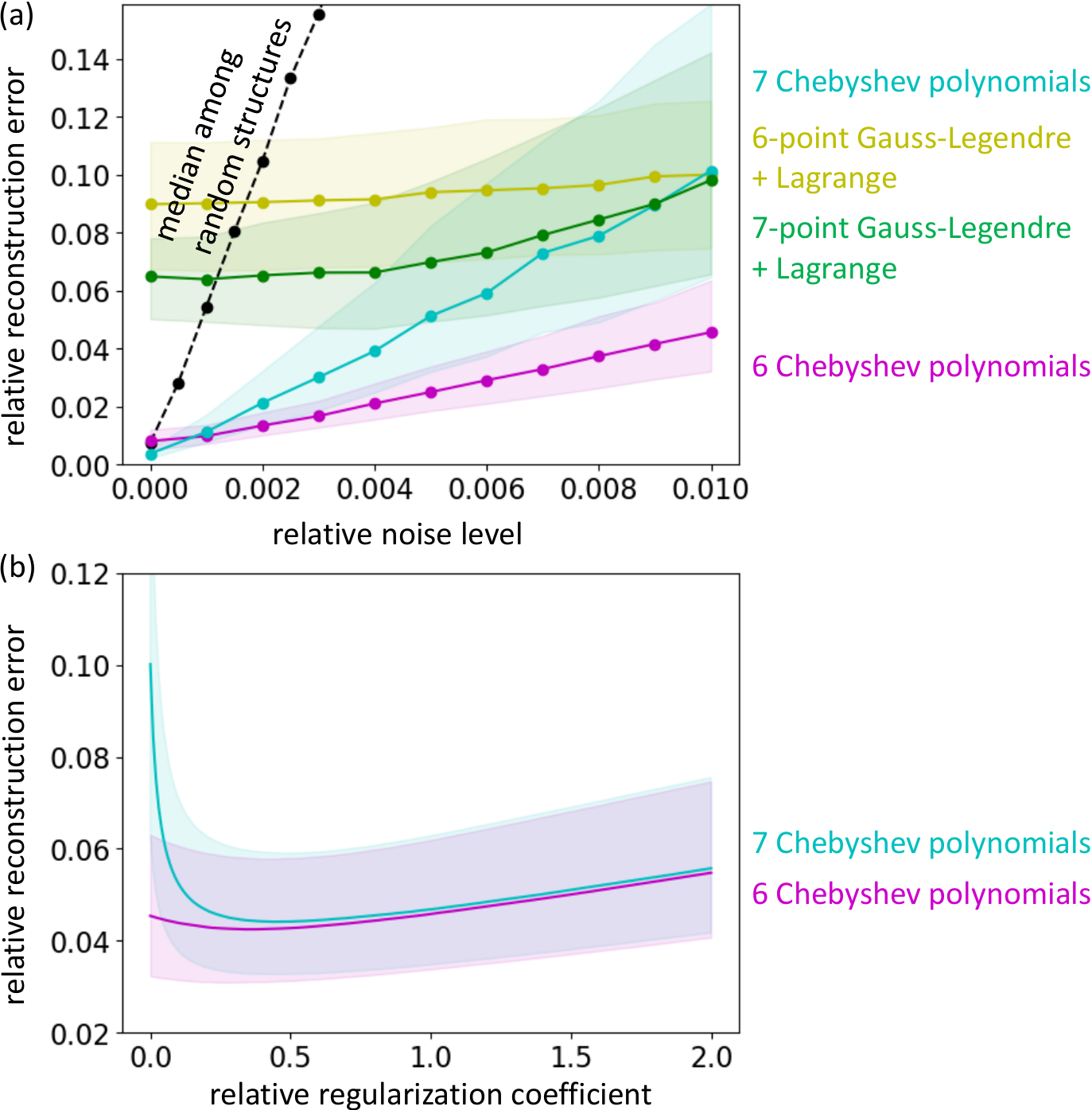}
\caption{(a) Dependence of reconstruction error on relative noise level. The horizontal and vertical coordinates represent the relative level $q$ of sensor noise in Eq.~(\ref{eq:noise}) and relative reconstruction error $\sqrt{\frac{1}{\omega_{\max}-\omega_{\min}}\int [\hat u(\omega)-u(\omega)]^2 d\omega}/\langle u(\omega)\rangle$ with $\langle u(\omega)\rangle=u_{\rm{avg}}$, where the integral is evaluated at 101 frequencies located at the (scaled) Gauss–Legendre quadrature nodes. Each dot represents the median among 2000 cases with randomly generated spectra and noise according to Eqs.~(\ref{eq:spectra}) and (\ref{eq:noise}). Each error band ranges from the first to third quartiles. The dots on the solid and dashed curves are based on the optimized and random structures, respectively. Here, the reconstruction errors for random structures are only for the reconstruction with 6 Chebyshev polynomials. The random structures, corresponding to the circular dots in Fig.~\ref{compare}, have the same minimum lengthscale as the optimized structure.
(b) Reconstruction errors with Tikhonov regularization. The horizontal coordinate represents a relative regularization coefficient, which we defined as $\alpha/\langle\sigma\rangle_{\rm geo}$ with $\langle\sigma\rangle_{\rm geo}$ denoting the geometric mean of the singular values of $F\sqrt{W}$ evaluated at the 7 Gauss–-Legendre quadrature nodes (scaled to the frequency range). Each curve represents the median among $10^5$ cases with randomly generated spectra and noise.
For the schemes with 7 and 6 Chebyshev polynomials at a relative noise level $q=0.01$, the optimal regularization coefficients $\alpha$, at which the median reconstruction errors attain their minima, are $0.85\langle\sigma\rangle_{\rm geo}$ and $0.71\langle\sigma\rangle_{\rm geo}$, respectively. Correspondingly, the ratios between the minimized reconstruction errors under regularization and the errors without regularization are $0.441$ and $0.936$.}\label{noise}
\end{figure}

Reconstruction errors generally increase with the noise level, as illustrated in Fig.~\ref{noise}(a). Here, the simulation is based on 2000 randomly generated spectra satisfying Eq.~(\ref{eq:spectra}), each of which suffers from sensor noises described by Eq.~(\ref{eq:noise}), and the noise in each spectrum is independent. We consider reconstruction from both 6 and 7 basis functions or interpolation points: changing the number of degrees of freedom, leading to well-known trade-off between accuracy and robustness.
Although 6-point reconstruction at Gauss–Legendre nodes does not exhibit advantages, reconstruction from 6 (zeroth- to fifth-order) Chebyshev polynomials appears to be more robust against sensor noise. Tikhonov regularization~\cite{Hansen1998} can decrease reconstruction error further. As Fig.~\ref{noise}(b) shows, for schemes with 7 and 6 Chebyshev polynomials, at the relative noise level $q=0.01$, with a properly chosen regularization coefficient $\alpha$, the median reconstruction errors decrease by 55.9\% and 6.4\%, respectively, from those without regularization. The simulation here is based on $10^5$ randomly generated spectra. We also explored reconstruction with Gaussian basis functions (a form of radial basis function~\cite{Buhmann2003}) similar to Refs.~\citenum{Yang2019,Cheng2021}, and found that they could obtain accuracy similar to the Chebyshev polynomials, but required careful tuning of the widths of Gaussians (see Sec.~S1 in Supporting Information).

For comparison, in Sec.~S4 in Supporting Information, we also investigated a more traditional end-to-end approach to the same problem, in which we directly minimized the mean reconstruction error $\sim \langle \Vert \hat{u} - u \Vert^2_2 \rangle$ over randomly sampled spectra and noise.  Although a similar noise-robustness may eventually be obtained, we found that the popular Adam algorithm~\cite{Kingma2014} for stochastic optimization converged much more slowly (exhibiting worse noise robustness if we terminated it at a number of Maxwell solves equal to our nuclear-norm FOM with CCSA), even with hyperparameter tuning.

%One may also choose unequally spaced frequency points, such as those in Gauss–Legendre quadrature. We applied the two choices as examples with six frequency points, and then performed cubic spline interpolation, as shown by the blue and green dots and curves in Figs.~\ref{fig4}(a)-(d).

%In the above examples, one can see that the Chebyshev reconstruction performs better. This trend persists under small noise, as shown in Fig.~\ref{fig4}(e).

\section{Concluding remarks}\label{sec:conclusion}

Although end-to-end co-design of optics and inference, incorporating training data and noise directly into the optimization process, continues to be an exciting area of research, we believe that this work illustrates new opportunities for devising inference-related figures of merit (FOMs) for inverse design decoupled from specific inference algorithms or training data.   We expect that identifying such figures of merit will be fruitful for many problems besides spectrometry, perhaps extending from polarimetry and imaging to object recognition and communications.  Moreover, such FOMs, which depend only on the optical properties of the system, may be promising vehicles for identifying, proving, and approaching theoretical upper bounds on the attainable performance.  Many such FOMs could potentially be explored, from linear-algebraic quantities such as norms and condition numbers (which have many possible variations), to quantities inspired more by information theory or entropy~\cite{Pinkard2025,Kabuli2025,Amaolo2025}.  We view such deterministic FOMs as a useful \emph{complement} to end-to-end co-design methods that directly minimize reconstruction error~\cite{Sitzmann2018,Lin2021,Lin2022,Li2023,Arya2024}: the former are more easily analyzed and may be more efficient to optimize, but the latter are more flexible in that end-to-end methods can be applied to more complicated reconstruction algorithms for which analytical FOMs are not yet known.

For the specific case of spectrometry, an important area of investigation is the optimal reconstruction of spectra that include both smooth background and sharp spikes (e.g.~absorption or emission lines); the latter should be amenable to sparse/compressed-sensing methods~\cite{Hastie2015}, but a combination of smooth and sparse methods is desirable for spectra containing both features.   To optimize underdetermined sparse reconstruction, one possibility is an end-to-end approach~\cite{Arya2024}. Although a quantity similar to ${\rm tr} (FF^\top)^{-1}$ has been suggested as a figure of merit for underdetermined reconstruction~\cite{Yu2025}, that quantity corresponds more closely to a minimum-norm prior via the pseudo-inverse~\cite{Strang2019} rather than a sparsity prior.
If the spectra of interest are even more restricted, characterized by a well-understood and extensive training set, more specialized data-driven reconstruction strategies become applicable, such as neural networks~\cite{Aggarwal2018}.  For much larger devices operating on wider bandwidths, one could first demultiplex the bandwidth into a set of narrow windows~\cite{Zheng2019,ZhangZ2022,Zhang2024} and then apply our methodology to spectrometry within each window using similar device footprints.   Although our designs in this paper were 2D, our objective function is also applicable to 3D with the cost dominated by the Maxwell solves, which have already been demonstrated for 3D inverse designs with similar footprints~\cite{Hammond2022,Yu2025}.

% We have introduced a method for inverse design of computational spectrometers, and demonstrated this method in integrated optical systems.
% Some extensions are possible. For example, one may raise the power of the matrix under the trace operation in Eq.~(\ref{eq:obj-singvals}) to the $n$-th power with $n>0$, resulting an objective function equal to the sum of all $\sigma_j^{-n}$.
% Further refinements of this approach may be worth investigation, for example, choosing the most suitable power $n$ to start with and adjusting it during optimization. An objective function that combines different $n$ or uses other matrix norms may also be possible. In addition, a spectrum could be expressed as a nonlinear function of unknown coefficients, such as some type of neural network. How to adapt our approach to these more general cases may also be of interest. Our method is not limited to spectrometry, but can also be applied in computational imaging, spectral imaging, and even beyond optics.

%More generally, one may use the nuclear norm of the $n$-th power of $(F\sqrt{W})^+$ with $n>0$. This objective function is equal to the sum of all $\sigma_j^{-n}$:
%\begin{equation}\label{eq:obj-singvals-n}
%\left\Vert\left[(F\sqrt{W})^+\right]^n\right\Vert_*={\rm tr}\left\{\left[(F^{\top}F)^{-1/2}W^{-1/2}\right]^n\right\}=\sum_j\frac{1}{\sigma_j^n}.
%\end{equation}

{\bf Research funding:} This work was supported in part by the US Army Research Office through the Institute for Soldier Nanotechnologies (Award No.~W911NF-23-2-0121) and by the Simons Foundation through the Simons Collaboration on Extreme Wave Phenomena Based on Symmetries. R.P. was supported in part by the National Institutes of Health (Award No.~R21EB036343).
Z.L. was supported in part by the US Army Research Office (Award No.~W911NF2410390) and Department of Energy (Award No.~DE-SC0024223).

{\bf Author contributions:} All authors have accepted responsibility for the entire content of this manuscript and approved its submission.

{\bf Conflict of interest:} Authors state no conflicts of interest.

{\bf Data availability:} The data that support the findings of this study are available from the corresponding author upon reasonable request.

\renewcommand{\thefigure}{S\arabic{figure}}
\renewcommand{\thesection}{S\arabic{section}}
\renewcommand{\theequation}{S\arabic{equation}}

\begin{center}
    \LARGE\bfseries Supporting Information
    \vspace{0.5em}
\end{center}

\setcounter{section}{0}
\setcounter{figure}{0}
\setcounter{equation}{0}

\section{Spectral reconstruction with 6 points or basis functions}
\begin{figure}[ht]
\centering
\includegraphics[width=1\linewidth]{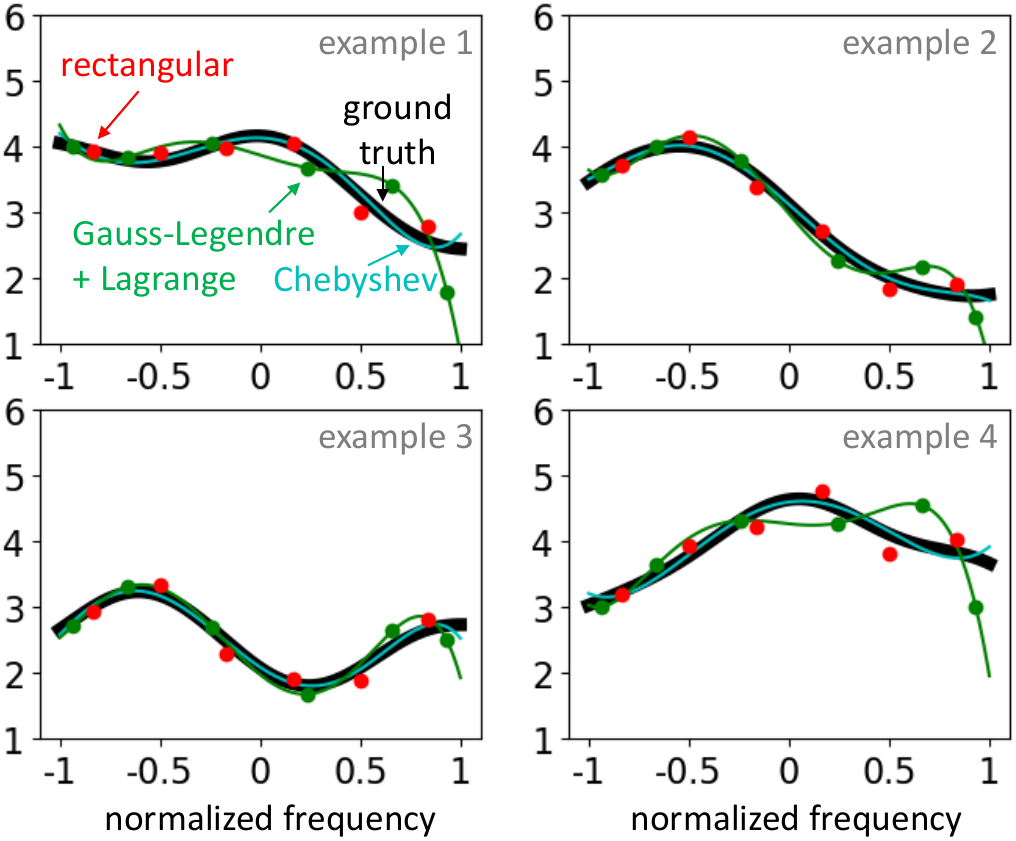}
\caption{Spectral reconstruction with four examples. The ground truth is plotted as thick black curves. The red dots represent reconstructed spectra at 6 equally spaced frequencies, chosen according the rectangular rule. The green dots represent reconstructed spectra at 6 frequencies chosen according to Gauss–Legendre nodes, while their Lagrange interpolating polynomials are plotted as green curves. The cyan curves represent reconstructed spectra as a linear combination of the first 6 Chebyshev polynomials of the first kind.}\label{reconstruct-6}
\end{figure}

Let us consider spectral reconstruction with 6 points or polynomials in the absence of sensor noise. As shown in Fig.~\ref{reconstruct-6}, the results are similar to those with 7 points or basis functions shown in Fig.~5 of the main text.

\begin{figure}[ht]
\centering
\includegraphics[width=0.8\linewidth]{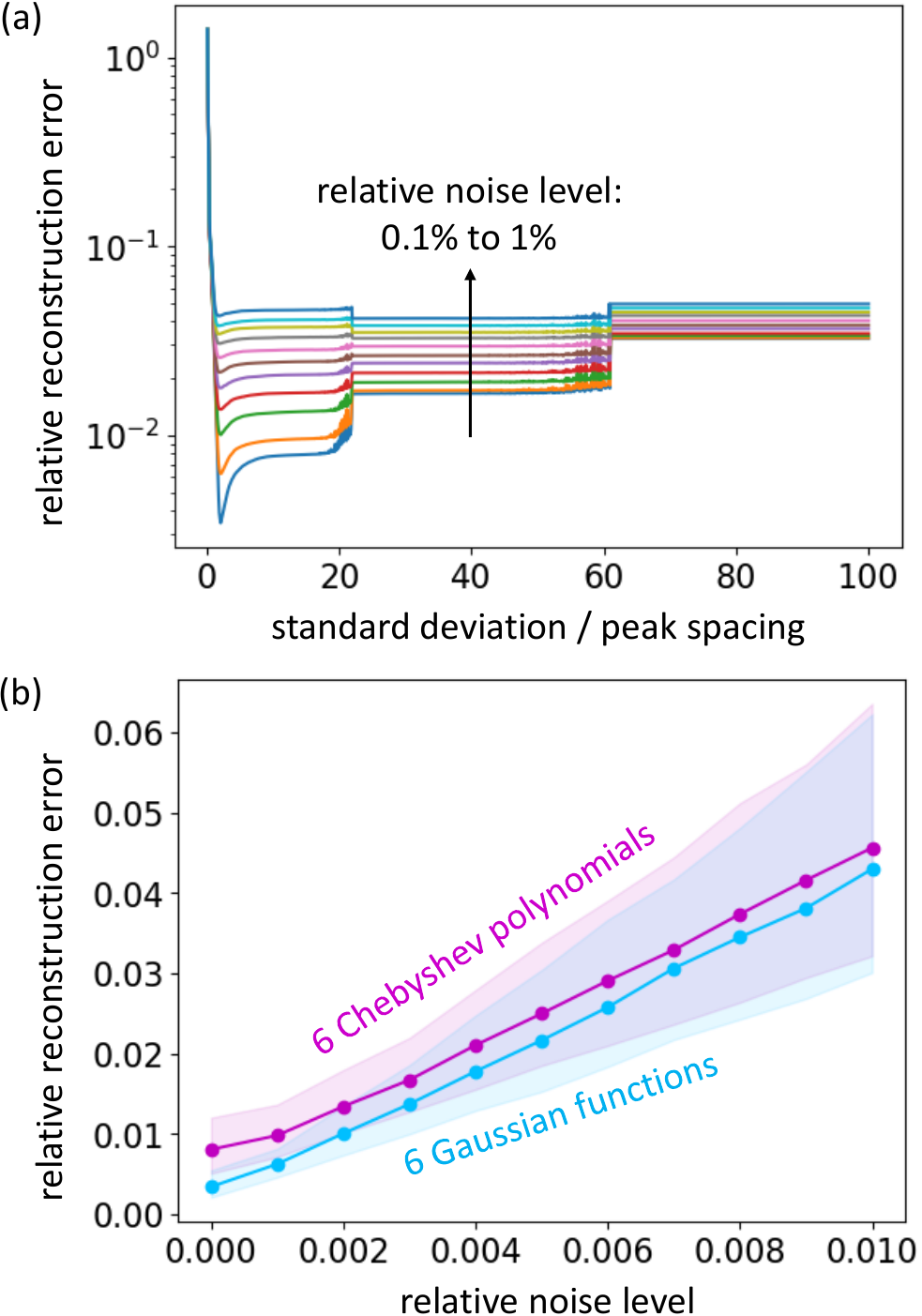}
\caption{Spectral reconstruction using 6 Gaussian functions, which are equally spaced and have an equal width. (a) Variation of relative reconstruction error with the standard deviation of the Gaussian functions under various relative noise levels ranging from $q=0.001$ to $0.01$. For $q=0.001$, the median relative reconstruction error is minimized at $\sigma=2.1\times$ peak spacing.
(b) Relative reconstruction error using this value of $\sigma$. The median relative reconstruction error and the range between the first and third quartiles are represented by the blue curve and shaded region. The relative reconstruction error using the 6 Chebyshev polynomials, the same as that in Fig.~6(a) of the main text, is also illustrated for comparison, as the magenta curve and shaded region.}\label{gaussian}
\end{figure}

Apart from polynomials, radial basis functions, such as Gaussian functions, are also a typical choice. Here, we consider 6 equally spaced Gaussian functions, with an equal width and their peaks located at the same frequencies as the red dots in Fig.~\ref{reconstruct-6}. As Fig.~\ref{gaussian}(a) shows, the relative reconstruction error, calculated in the same manner as those in Fig.~6(a) of the main text, varies with the noise level and the standard deviation $\sigma$ (proportional to the width) of the Gaussian functions. For the relative noise level $q=0.001$, the median relative reconstruction error is minimized at $\sigma=2.1\times$ peak spacing. At this value of $\sigma$, the variation of median relative reconstruction error with noise is illustrated as the blue curve in Fig.~\ref{gaussian}(b), with the shaded region representing the range between the first and third quartiles. For comparison, the relative reconstruction error using the first 6 Chebyshev polynomials of the first kind is also illustrated here as the magenta curve and shaded region, which is identical to those in Fig.~6(a) of the main text. 

\section{Influence of fabrication imperfections}
\begin{figure}[ht]
\centering
\includegraphics[width=0.68\linewidth]{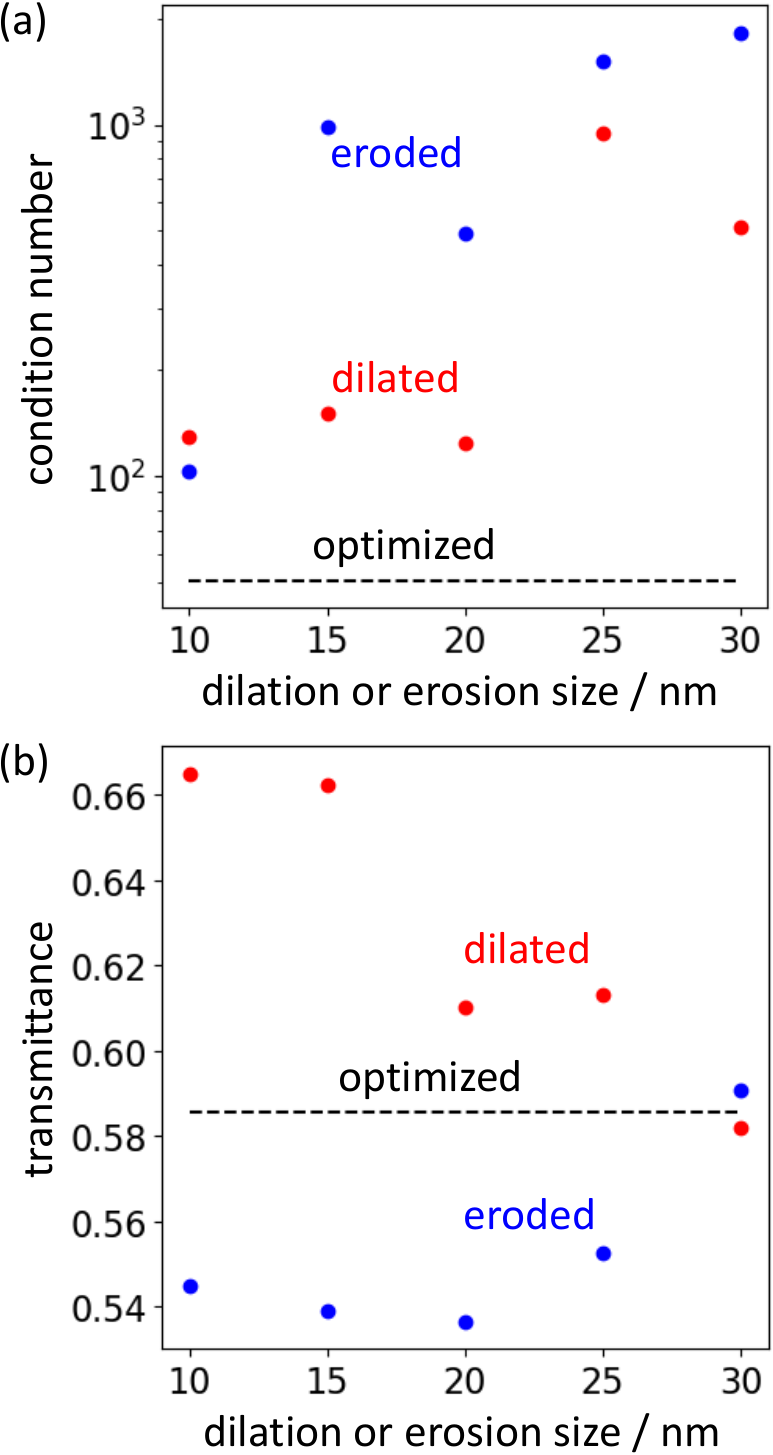}
\caption{Performance metrics under morphological dilation and erosion of the optimized design pattern. The red and blue dots represent the outcomes for dilated and eroded structures, respectively. The dashed lines label the performance metrics of the optimized design.}\label{dilation-erosion}
\end{figure}

\begin{figure}[ht]
\centering
\includegraphics[width=0.8\linewidth]{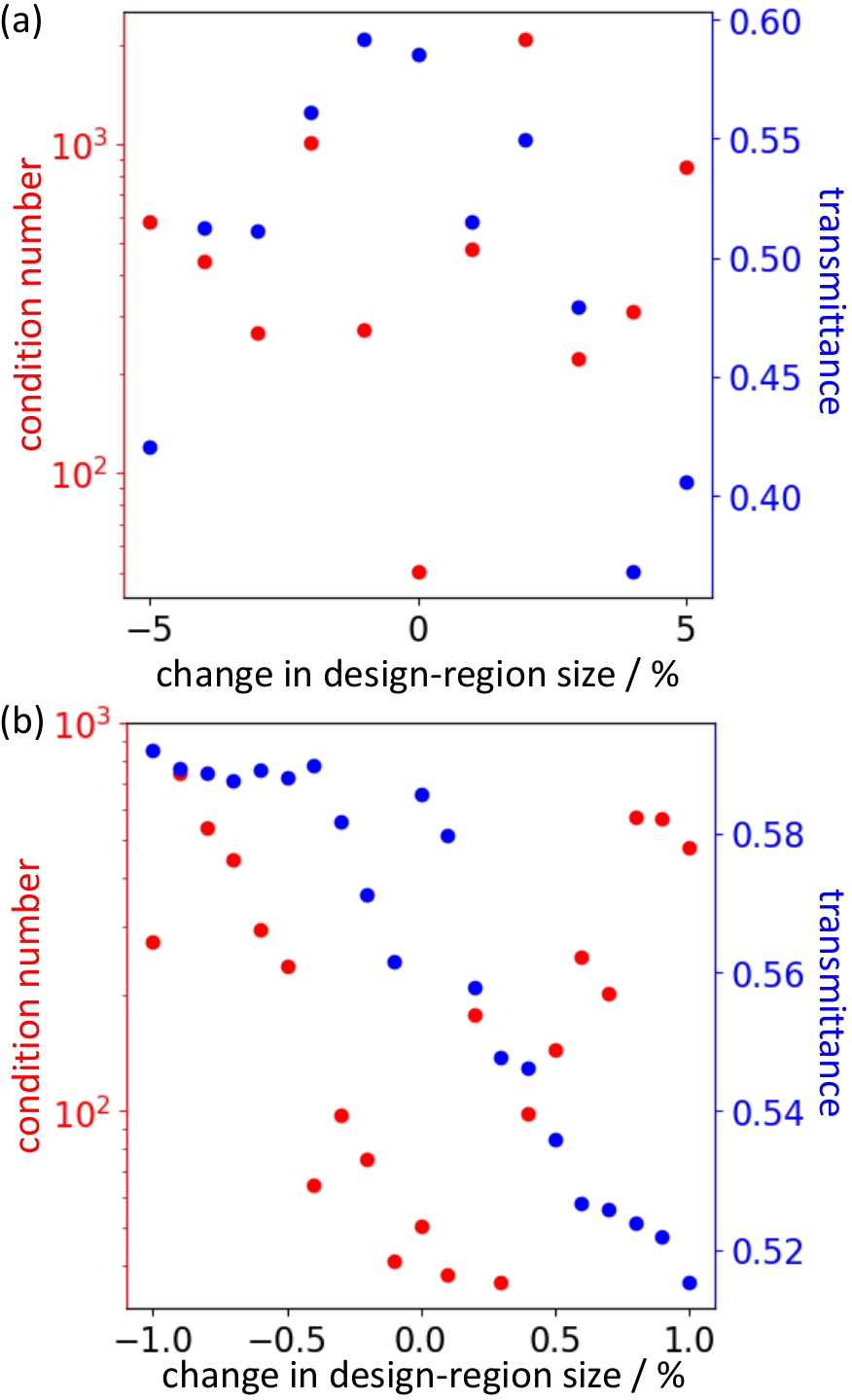}
\caption{Performance metrics under errors in the design-region size.}\label{size-change}
\end{figure}

In this section, for the optimized design in Fig.~3 of the main text, we show the performance metrics under imperfections in fabrication. The influences of different extents of dilation and erosion on the optimized design pattern are illustrated in Fig.~\ref{dilation-erosion}, while the influences of the size of the design region are shown in Fig.~\ref{size-change}. One can see that the condition number is more vulnerable to the fabrication errors compared with transmittance. The large fabrication imperfections ($>$ 10 nm) simulated here are for demonstration purposes and are not typical of realistic fabrication~\cite{Siew2021,Boning2022}.

\section{Analysis of objective functions}
As introduced in the main text, a general version of our objective function to be minimized is
\begin{equation}\label{eq:general-obj}
f(\rho) = {\rm tr}\left\{[(WF^{\top}F)^{-1}]^{n/2}\right\}=\sum_j{\sigma_j}^{-n},
\end{equation}
where $\rho$ represents design variables; the matrix $F$ and singular values $\sigma_j$ are also functions of $\rho$ but the dependence is not written explicitly; the exponent $n$ should be positive, with $n=1$ and $n=2$ resulting in our objective function in Eq.~(14) and the Fisher-information--inspired objective function, respectively. The function in Eq.~(\ref{eq:general-obj}) is dominated by the term with the smallest singular value. The gradient of the objective with respect to the design variables is
\begin{equation}\label{eq:general-obj-grad}
\nabla_{\rho} f =-n\sum_j\frac{\nabla_{\rho}\sigma_j}{{\sigma_j}^{n+1}},
\end{equation}
which implies that the variation in the smallest singular value also dominates the variation of the objective. The dependence is larger for large $n$.
The log-determinant objective and its gradient
\begin{equation}\label{eq:logdet}
\begin{aligned}
&g(\rho) = -\ln\det (WF^{\top}F)=-2\sum_j\ln\sigma_j,\\
&\nabla_{\rho}g = -2\sum_j\frac{\nabla_{\rho}\sigma_j}{\sigma_j},
\end{aligned}
\end{equation}
also have similar dependence on the smallest singular value and its change.

\section{Comparison with end-to-end design}
In this section, we compare the performance of our design method with end-to-end approaches and different optimization algorithms. To perform end-to-end design, a reconstruction scheme needs to be specified. Here, we consider the scheme described by Eq.~(7) of the main text, which is duplicated here---with basis functions discretized as $B$ and Tikhonov regularization, the reconstructed spectrum $\hat u$ can be computed as
\begin{equation}\label{eq:c-reconstr}
\begin{aligned}
&\hat c = \left[B^\top(WF^\top FW+\alpha W)B\right]^{-1}(FWB)^\top v,\\
&\hat u = B\hat c.
\end{aligned}
\end{equation}
In a typical end-to-end approach, the permittivity $\epsilon$ of the scatterer (which determines $F$), and possibly the regularization coefficient $\alpha$, are optimized to reduce the relative reconstruction error (backpropagating gradients ``end-to-end'' through both the physics and reconstruction algorithm):
\begin{equation}\label{eq:e2e-obj}
\min_{\epsilon, \alpha}\, \biggl\langle\frac{\left\Vert\hat u-u\right\Vert_2^2}{\left\Vert u\right\Vert_2^2}\biggr\rangle,
\end{equation}
where the angle brackets denote averaging over spectra and noise. Ideally, an end-to-end approach would involve a denser set of frequencies to compute the measured noisy-sensor readings accurately. Here, to lower computational cost, we selected 7 frequencies according to Gauss--Legendre quadrature as an approximation. 
End-to-end design is typically performed with stochastic optimization, in which the average in Eq.~(\ref{eq:e2e-obj}) is over the minibatch in each iteration.
Here, we employed the Adam algorithm with default hyperparameters~\cite{Kingma2014}. (Attempts with non-default hyperparameters did not achieve noticeably better performance in our tests.)

% We also experimented with an \emph{ad-hoc} optimization algorithm in which we applied the deterministic CCSA-MMA for 40 iterations at a time before resampling; although this is nonstandard and has not been analyzed in the literature, it worked well on this problem.

\begin{figure}[ht]
\centering
\includegraphics[width=0.8\linewidth]{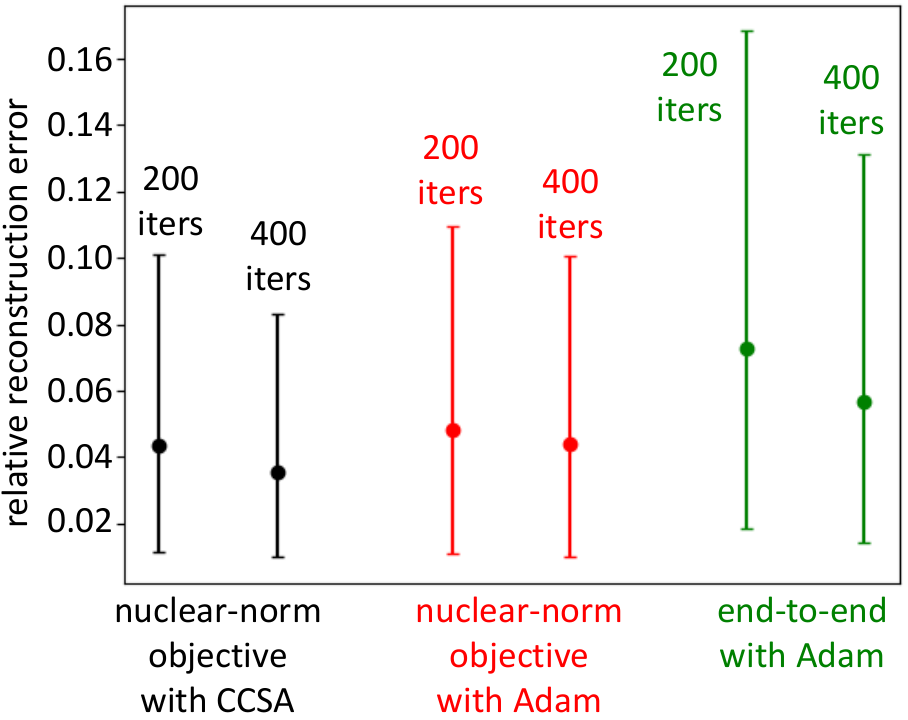}
\caption{Reconstruction errors associated with spectrometers designed by different methods. The relative reconstruction error denoted by the vertical axis is computed as $\sqrt{\frac{1}{\omega_{\max}-\omega_{\min}}\int [\hat u(\omega)-u(\omega)]^2 d\omega}/\langle u(\omega)\rangle$ with $\langle u(\omega)\rangle=u_{\rm{avg}}$, which is the same as that in Fig.~6 of the main text. The integral is evaluated at 101 frequencies located at the (scaled) Gauss–Legendre quadrature nodes. Each dot represents the median among $10^5$ cases with randomly generated spectra and noise according to Eqs. (16) and (17) in the main text. Each error bar ranges from the first to third quartiles.
The examples here are based on the same integrated structure as that sketched in Fig.~3(a) of the main text, but with the hyperparameter $\beta$ fixed at 8.
%The design from each method was optimized with 200 iterations.
The regularization coefficient $\alpha$ was optimized (or reoptimized if the end-to-end approach is used) after the design was given.}\label{end-to-end}
\end{figure}

%In Fig.~\ref{end-to-end}, the end-to-end approaches with minibatch sizes 1000 and 1 are compared with our inverse-design approach, in which the trace objective is minimized with the CCSA-MMA algorithm~\cite{Svanberg2002}. We also minimized the trace objective with our objective using the Adam algorithm. In these examples, the trace objective in our method objective results in faster convergence than the end-to-end approaches.
In Fig.~\ref{end-to-end}, the relative reconstruction error from an end-to-end approach (green) with a minibatch size 1000 is compared with our nuclear-norm inverse-design approach (black), in which the trace objective is minimized with the CCSA-MMA algorithm~\cite{Svanberg2002}. We also minimized our nuclear-norm FOM using the Adam algorithm (red), which turned out to have slower (but more monotonic) convergence than CCSA. The noise-robustness (related to the relative reconstruction error for noisy measurements) and transmittance of the resulting design are evaluated in Fig.~\ref{end-to-end} and Table~\ref{tab:end-to-end} for both 200 and 400 optimizer iterations (Maxwell solves).  In these examples, the nuclear-norm FOM with CCSA results in better performance than either the Adam-based end-to-end approach or the Adam algorithm applied to our deterministic FOM.  Although it is possible that running the Adam algorithm for long enough may eventually match the performance of the other methods, it seems to be converging more slowly---this is not too surprising, since Adam has to handle more challenging stochastic (non-deterministic) objectives.  It may be that with sufficient \emph{ad-hoc} algorithmic tuning one could improve the convergence rate of an end-to-end scheme, but we believe that greater flexibility in choosing optimization algorithms is an advantage of a determistic FOM.

% As an \emph{ad hoc} end-to-end experiment, we also tried the CCSA-MMA algorithm with a minibatch size 1000, but re-sampled the data and noise only every 40 iterations.  In this case our objective was the end-to-end reconstruction error $\langle(F^+\zeta)^{\top}W^{-1}(F^+\zeta)\rangle$, inspired by Eqs.~(8) and (9), where $\zeta$ denotes the column vector for sensor noise and the angle brackets denote averaging over noise. This expression being minimized approximates the naïve deviation $\langle\left\Vert\hat u-u\right\Vert_2^2\rangle$ with $\hat u=(FW)^+v$. We used a minibatch size 1000 with the data set for noise updated every 40 iterations, and employed the CCSA-MMA algorithm. As shown in Fig.~\ref{end-to-end} (blue), this atypical approach results in performance comparable to our nuclear-norm method. 

\begin{table}[ht]
\caption{Transmittance of spectrometers optimized with different methods and numbers of iterations}
\label{tab:end-to-end}
\begin{tabular}{c|c|c|c}
    & \makecell{nuclear-norm \\ objective \\ with CCSA}
    & \makecell{nuclear-norm \\ objective \\ with Adam}
    & \makecell{end-to-end \\ with Adam} \\ \hline
200 iterations & 71.5\% & 48.5\% & 50.3\% \\ \hline
400 iterations & 76.4\% & 53.7\% & 56.3\%
\end{tabular}
\end{table}

\section{Integrated computational spectrometer without output waveguides}\label{sec:n2f}
%\subsection{Example structure}
\begin{figure*}[ht]
\centering
%\fbox{\includegraphics[width=0.7\linewidth]{sketch}}
\includegraphics[width=0.65\linewidth]{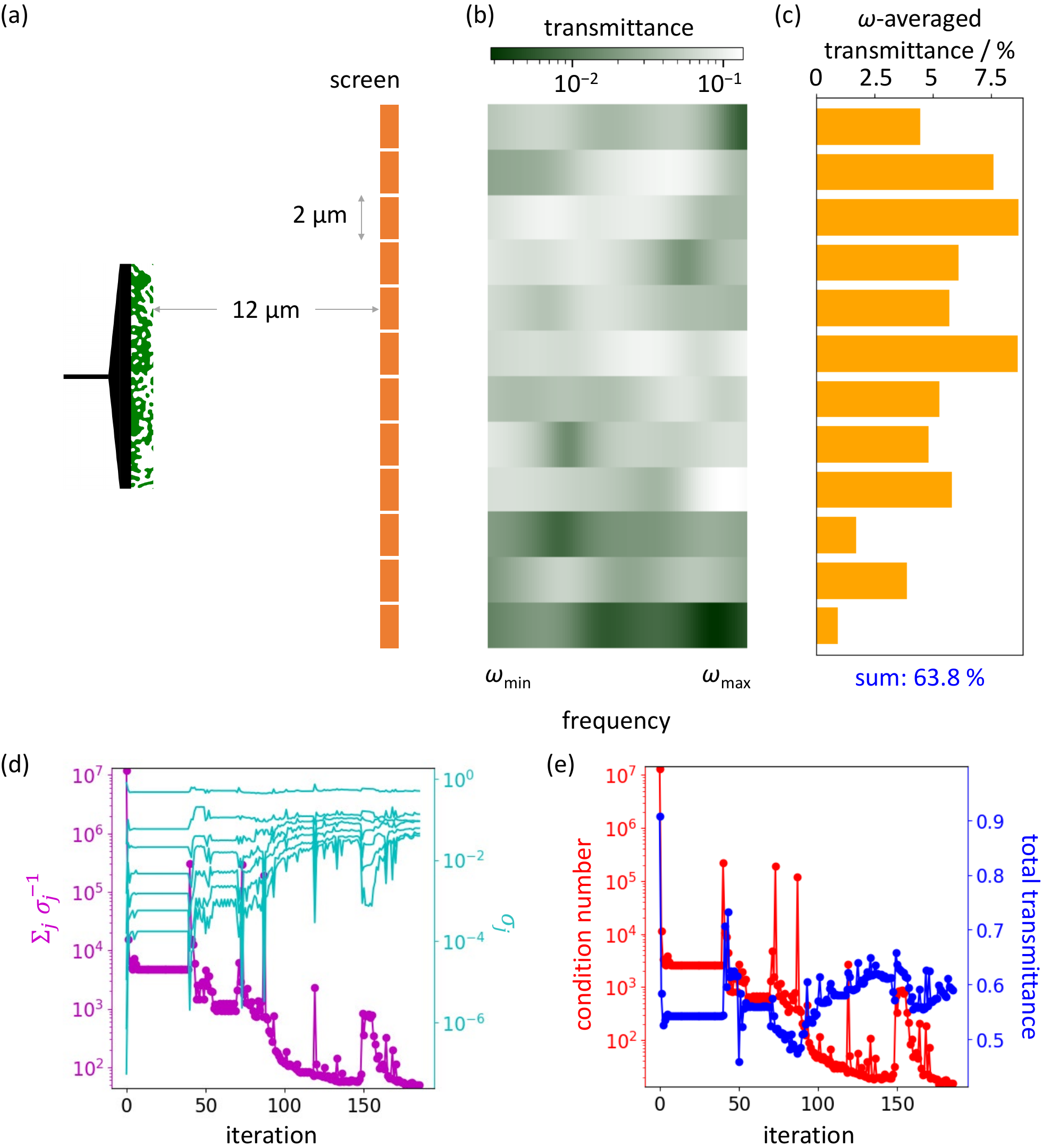}
\caption{Inverse design of an integrated computational spectrometer without output waveguides. (a) Structure of the spectrometer. This 2d structure consists of an input waveguide, a wedge region, a design region, and a screen equipped with 12 sensors, with the solid material having a relative permittivity $\approx12$. The sizes of each sensor and the screen are 2 {\textmu}m and 24 {\textmu}m, respectively. Apart from the free space and screen in place of the output waveguides, the other components are the same as those in Fig.~3(a) of the main text.
(b) Transmittance of the optimized spectrometer at each output waveguide across the frequency range of interest.
(c) Frequency-averaged transmittance of the optimized spectrometer at each output waveguide. The total transmittance is 63.8\%.
(d) Objective function ($\sum_j\sigma_j^{-1}$) and singular values during optimization.
(e) Condition number and total transmittance during optimization, computed from the $12\times7$ spectral--spatial mapping matrix.}\label{optimize-n2f}
\end{figure*}

Here, we demonstrate our methods on a simple two-dimensional (2d, $xy$) example of an integrated spectrometer, in which output signals are collected by sensors on a screen distant from the scatterer, as Fig.~\ref{optimize-n2f}(a) shows. The input waveguide, the wedge region, and the design region have the same sizes and relative positions as those in Fig.~3(a) of the main text. The output waveguides are replaced by free space, while 12 sensors, each with a size of 2 {\textmu}m, are closely arranged on a screen parallel to the design region with a distance of 12 {\textmu}m, as depicted in Fig.~\ref{optimize-n2f}(a). The signal recorded by each sensor is proportional to the power of waves traveling through that sensor. Incoming waves have wavelengths between 1.54 and 1.56 {\textmu}m with out-of-plane ($E_z$) polarization, which are the same as those in the main text. We selected seven frequencies according to Gauss--Legendre quadrature of $\int_{1.54}^{1.56}d\lambda$ for performing inverse design, so the size of the spectral--spatial mapping matrix is again $12\times7$.

\begin{figure*}[ht]
\centering
%\fbox{\includegraphics[width=0.7\linewidth]{sketch}}
\includegraphics[width=0.55\linewidth]{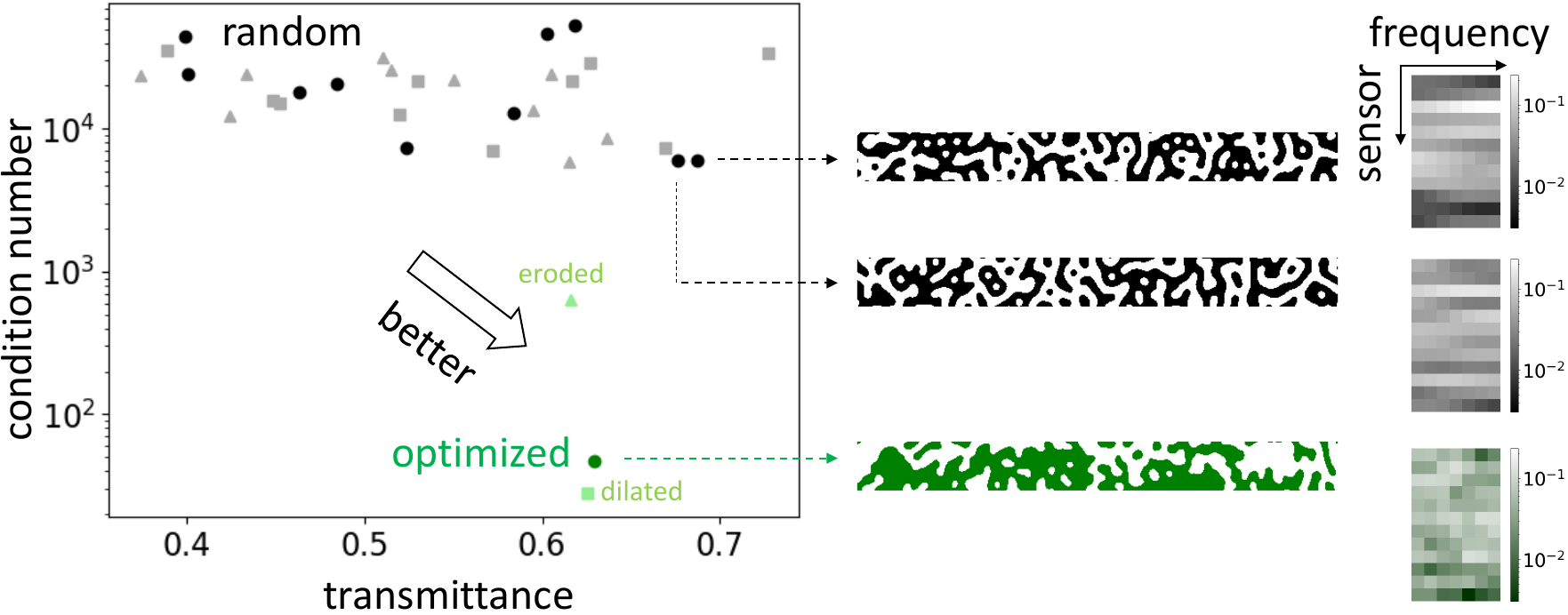}
\caption{Comparison of performances of random and optimized structures. The horizontal and vertical axes represent the total transmittance and the condition number, respectively. The bottom right area is associated with lower condition numbers, higher total transmittance, and hence better performance. The round green and black dots correspond to the optimized and random structures with the same minimum lengthscale, while the square/triangular symbols with light colors correspond to structures in which solid regions are dilated/eroded by 10 nm. The optimized structure and two random structures are illustrated on the middle panel. Their corresponding $12\times7$ spectral--spatial mapping matrices are on the right panel.}\label{compare-n2f}
\end{figure*}

As shown in Fig.~\ref{optimize-n2f}, the process and result of inverse design are similar to those in Fig.~3(a) of the main text. However, in comparison with random structures with the same minimum lengthscale (80 nm), the optimized design has moderate total transmittance but much lower condition number, as Fig.~\ref{compare-n2f} illustrates.

\section{FDTD simulation for optimization}
All electromagnetic simulations in optimization and verification were performed with a free
and open-source implementation of the finite-difference
time-domain (FDTD) method~\cite{Oskooi2010}. The inverse design and end-to-end design were performed with the hybrid time/frequency-domain
adjoint module of Meep. In the design process, the size of the simulation cell is 14 µm × 6 µm, whose periphery is a perfectly matched layer (PML) with a thickness of 1 µm. The resolution of FDTD simulation is 50 pixels/µm, while the resolution of the material grid, which is separate from the simulation resolution, is 200 pixels/µm. Seven frequencies are involved.
%In the design process the simulation resolution is 50 pixels/µm, while in verification the simulation resolution is 200 pixels/µm. The resolution of the material grid, which is separate from the simulation resolution, is always 200 pixels/µm.

Each iteration in optimization is composed of a forward simulation and an adjoint simulation. On a supercomputing node with 48 CPUs (Intel Xeon
Platinum 8260), an iteration in designing the spectrometer in the main text takes 200 to 300 sec, while an iteration in designing the spectrometer in Sec.~\ref{sec:n2f} takes 700 to 800 sec. The design processes in the two cases undergo roughly 300 or 200 iterations, which take about one or two days in total.

~\\
~\\
~\\
~\\
~\\
~\\
~\\
~\\
~\\
~\\
~\\
~\\
~\\
~\\
~\\
~\\
~\\
~\\
~\\
~\\
~\\
~\\
~\\
~\\
~\\
~\\
~\\
~\\
~\\
~\\
~\\
~\\
%\bibliographystyle{unsrt}
%\bibliography{main}
\providecommand{\noopsort}[1]{}\providecommand{\singleletter}[1]{#1}%

\end{document}